\documentclass[12pt]{article}
\usepackage{amsmath}
\usepackage{graphicx}
\usepackage{enumerate}
\usepackage{bbm}
\usepackage{natbib}
\usepackage{dsfont}
\usepackage{url} 
\usepackage{relsize,color}
\usepackage[dvipsnames]{xcolor}
\usepackage{setspace}
\usepackage{etoolbox}
\AtBeginEnvironment{tabular}{\onehalfspacing}
\usepackage{multirow}

\usepackage{algpseudocode}
\usepackage{algorithm}
\usepackage{epsfig}
\usepackage{ulem} 
\usepackage{xspace} 
\usepackage{graphicx}
\usepackage{sidecap}
\usepackage{times,epsfig,makeidx,amsfonts,amsmath}
\usepackage{lscape} 
\graphicspath{{figures/}} 

\newcommand{\E}{\text{E}} 

\newcommand{\di}{{\rm d}}

\newcommand{\bfs}{\textbf{s}}

\newcommand{\bfn}{\textbf{n}}

\newcommand{\bft}{\textbf{t}}

\newcommand{\lc}[2]{\sout{#1}\textcolor{blue}{\xspace#2}} 
\usepackage{ulem}

\newcommand{\blind}{1}

\begin{document}

\def\spacingset#1{\renewcommand{\baselinestretch}%
{#1}\small\normalsize} \spacingset{1}

	\vspace{-5in}
\if1\blind
{

  \title{\bf Adaptive preferential sampling in phylodynamics}
  \author{Lorenzo Cappello$^1$, Julia A. Palacios$^{1,2}$\thanks{ JAP acknowledges support from National Institutes of Health grant
  		R01-GM-131404 and the Alfred P. Sloan Foundation.}\\
    $^1$Department of Statistics, Stanford University\\
$^2$Department of Biomedical Data Science, Stanford Medicine}
  \maketitle
} \fi

\if0\blind
{
} \fi

\begin{abstract}
	Longitudinal molecular data of rapidly evolving viruses and pathogens provide information about disease spread and complement traditional surveillance approaches based on case count data. The coalescent is used to model the genealogy that represents the sample ancestral relationships. The basic assumption is that coalescent events occur at a rate inversely proportional to the effective population size $N_{e}(t)$, a time-varying measure of genetic diversity. 
	When the sampling process (collection of samples over time) depends on $N_{e}(t)$, the coalescent and the sampling processes can be jointly modeled to improve estimation of $N_{e}(t)$. Failing to do so can lead to bias due to model misspecification. However, the way that the sampling process depends on the effective population size may vary over time. We introduce an approach where the sampling process is modeled as an inhomogeneous Poisson process with rate equal to the product of $N_{e}(t)$ and a time-varying coefficient, making minimal assumptions on their functional shapes via Markov random field priors. We provide scalable algorithms for inference, show the model performance vis-a-vis alternative methods in a simulation study, and apply our model to SARS-CoV-2 sequences from Los Angeles and Santa Clara counties. The methodology is implemented and available in the  \texttt{R} package \texttt{adapref}.
\end{abstract}

\noindent%
\vfill

\spacingset{1.5} 
\section{Introduction}
\label{sec:intro}

Molecular sequence data, within the framework of phylodynamics \citep{gre04}, is increasingly being used to track disease spread caused by rapidly evolving viruses and pathogens such as Influenza viruses \citep{ram08},  Zika \citep{far16}, and SARS-CoV-2 \citep{had18}. The coalescent process \citep{king82,kingn82}, a probability model of gene genealogies, depends on a parameter called effective population size $N_e(t)$, which is a time-varying measure of genetic diversity. When disease dynamics can be modeled by simple epidemiological models such as Susceptible-Infected-Recovered, the coalescent effective population size can be expressed in terms of transmission rates and prevalence \citep{vol09,fro10}. Accurate and scalable inference for $N_e(t)$ is thus relevant to estimate epidemiological parameters of great interest in public health.
Although this work is motivated by applications in molecular epidemiology of infectious diseases, estimation of $N_e(t)$ is an active area of research with applications ranging across many other scientific domains such as conservation biology and population genetics (\textit{e.g.} \cite{sha04,huf10,lor11}).

A common feature in these applications is that genetic data are collected sequentially (heterochronous samples). In viral studies, samples are collected and sequenced when infected individuals attend clinics, hospitals, or testing centers. In ancient DNA studies, specimens are dated according to the time they lived, estimated through radiocarbon dating or other techniques. The coalescent typically models the gene genealogy conditionally on sampling dates, that is, the sampling dates are treated as censoring information \citep{rod99}. However, in some situations, it is reasonable to assume that samples are collected at a higher frequency when the population is large and at a lower frequency when the population is small: for example, at the onset of an epidemic, as the viral population grows and more people get infected, more resources may be allocated to monitor the viral spread, possibly leading to more molecular sequence collection. The number of SARS-CoV-2 sequences uploaded daily in GISAID offers some evidence of this claim \citep{shu2017gisaid} (see the histogram in the supplementary material). 

\cite{kar16} study the scenario in which the sampling process depends on the population size, and show that an estimator of $N_e(t)$ that does not account for this dependence is biased. This issue was first discussed in the spatial statistics literature by \cite{dig10}, who term preferential sampling a situation in which the process that determines the data locations and the process under study are dependent. In this paper, we will introduce a new model that account for preferential sampling in a coalescent framework, while making minimal assumptions on $N_e(t)$, the sampling process, and their dependence.  

Three estimators that incorporate preferential sampling into the coalescent framework have been proposed. \cite{vol14} propose an estimator in the case that $N_e(t)$ grows exponentially and samples are collected as an inhomogeneous Poisson process with rate linearly dependent on the effective population size.  \cite{kar16} assume that $N_e(t)$ is a continuous function, and the samples are collected as an inhomogeneous Poisson process with rate $\lambda(t)=\exp(\beta_0)N_e(t)^{\beta_1}$, for $\beta_0,\beta_1\geq0$, i.e. the dependence between the sampling process and the effective sample size is described by a parametric model. 
Recent work by \cite{par20} weakens this assumption substantially, allowing for the dependence between the sampling process and effective population size to vary over time. The key assumption in \cite{par20} is that the sampling rate depends linearly on $N_e(t)$ within a given time interval, but the linear coefficient changes across time intervals. The estimator of \cite{par20}, termed Epoch skyline plot (ESP), is an extension of the classic skyline plot estimator for $N_e(t)$ \citep{py00}, in which the sampling rate and the effective population size are both piecewise-constant, and the location and number of change points  (boundary points of the time intervals) are either specified or inferred. As it is typical with skyline plots, the estimates are highly variable, rough, and highly dependent on the specification of change points locations and the number of piecewise-constant pieces used. All three works show that under correct model specification, accounting for preferential sampling leads to a more accurate estimation of $N_e(t)$ (in terms of absolute deviations to the true trajectory), and narrower credible regions.

Other non-coalescent approaches for phylodynamics, such as birth-death processes \citep{stad10}, explicitly incorporate the sampling process by modeling the sampling dates as a partially observed death process where only a fraction of the population is observed. \cite{sta13} extended previous work to allow sampling rates to vary through time. \cite{vol14} show that in both, coalescent and birth-death processes alike, statistical power largely depends on the correct specification of the sampling process rate, rather than on the genealogical model. Hence, the need for a flexible modeling approach of the sampling process, adaptive to any possible scenarios encountered in applications. 


There are a plethora of nonparametric estimators of $N_e(t)$ following the skyline plot \citep{py00}: among others, the generalized skyline plot \citep{stri01} and the Bayesian skyline plot \citep{dru05} reduce the high variance and roughness that characterize the skyline plot estimators. These methodologies require either fixing or estimating change-points in $N_{e}(t)$. A set of models that do not employ change-points but arbitrary discrete grids is based on Markov random fields (MRF): the Gaussian MRF (GMRF) \citep{min08,pal12} allows for the recovery of smooth continuous trajectories; the Horseshoe MRF (HSMRF) \citep{fau20}  is an alternative to GMRF which is locally adaptive, \textit{i.e.} it can successfully recover sharp changes in a trajectory and it is adaptive to a varying level of smoothness.


In this paper, we borrow from this literature and introduce an adaptive preferential sampling framework for phylodynamics, where the adaptivity follows from the fact that the dependence of the sampling process on $N_e(t)$ changes over time. The effective population size $N_e(t)$ is modeled as a latent parameter included in both, the coalescent and the sampling processes. The latter assumed to be an inhomogeneous Poisson process with rate $\lambda(t)=\beta(t) N_e(t)$, where $\beta(t)$ is a continuous function controlling the dependence on $N_e(t)$, analogous to that introduced by \cite{par20}. We \textit{a priori} model $N_e(t)$ and $\beta(t)$ as two Markov random fields (MRF), with the flexibility of using either a GMRF or a HSMRF. The prior choice follows from the properties of the fields. The advantage of the proposed adaptive preferential sampling over the ESP estimator is that there is no need to specify (or estimate) the number and location of the change points of $\beta$ and $N_e$. Also, the resulting estimates are smooth and the high variability that characterizes skyline estimates disappears.

We develop the methodology assuming that a genealogy is available to the researcher and develop algorithms for inference under this framework. We test our model on simulated data and compare it to alternative methods, including both, estimators that account for preferential sampling and others that do not. We implement our method in the \texttt{R} package \texttt{adapref}, available at \texttt{ https://github.com/lorenzocapp/adapref}, provide two algorithms for posterior approximation: a  Hamiltonian MCMC and a Laplace approximation. We apply our method to SARS-CoV-2 sequences from California and study whether there is evidence of preferential sampling. 

The rest of the paper proceeds as follows. In Section 2, we provide background on the coalescent process, the MRF priors on $N_e(t)$, and previous work on preferential sampling. In Section 3 we introduce the adaptive preferential sampling framework and explain how to approximate the posterior distribution of model parameters. Section 4 includes a simulation study, in which we test the new set of priors through simulated data and compare them to alternatives. In Section 5, we apply our method to two data sets of SARS-CoV-2 sequences from Santa Clara and Los Angeles counties in California. Section 6 concludes. 

\section{Background}

\subsection{Coalescent model}
\label{sec:coal}

\begin{SCfigure}[2][!t]
	\centering
	\includegraphics[scale=0.65]{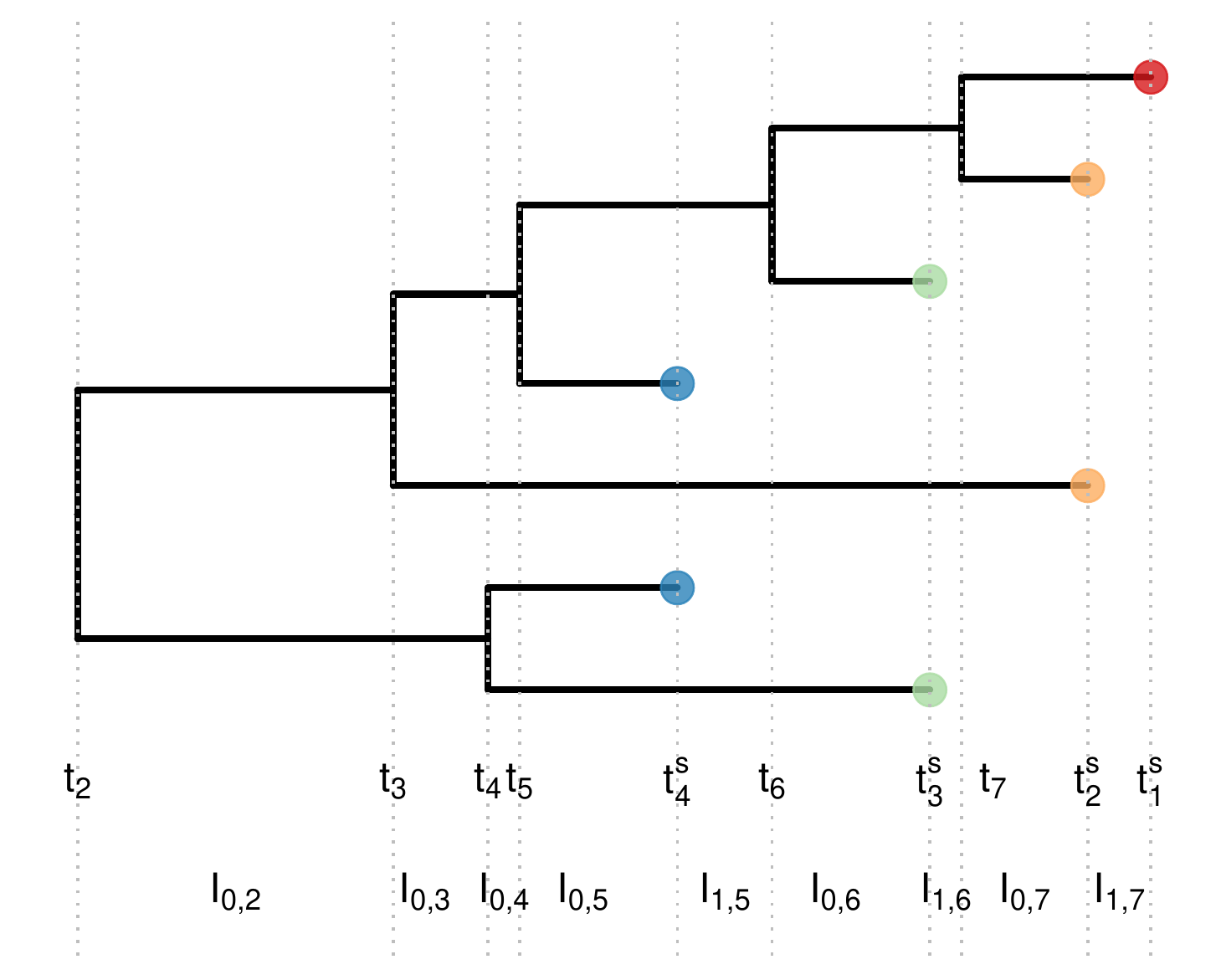}
	\vspace{-1cm}
	\caption{\small{\textbf{Example of a heterochronous genealogy.} A genealogy of $7$ individuals sampled at 4 different  times (color of tips) with multiplicities $(n_1=1,n_2=2,n_3=2,n_4=2)$}. Sampling times are denoted by $(t^{s}_{k})_{1:4}$, coalescent times are denoted by $(t_{k})_{2:7}$ and $I_{i,j}$ denoted the interval lengths delimited by coalescent times and/or sampling times, i.e. every time there is a change in the number of lineages.}
	\label{fig:het_tree}
\end{SCfigure}

Coalescent models are continuous-time Markov chains used to model the set of ancestral relationships of a sample of $n$ individuals from a large population called gene genealogy. In the context of molecular epidemiology, a genealogy is a subset of the transmission history among the samples (Figure~\ref{fig:het_tree}). Starting from the original work of \cite{kingn82}, several extensions to the standard coalescent have been developed to incorporate more realistic population and sampling features, such as variable population size \citep{sla91}, longitudinal sampling (also called heterochronous sampling) \citep{rod99} and population structure \citep{Hudson1990}; \cite{wak09} provides a good introduction to the subject. 
Coalescent processes can be characterized by two underlying processes: a jump chain defining the ancestral relationships represented by a binary tree topology and a pure death process that defines the timing of the coalescent events, i.e. the times when pairs of lineages meet their common ancestors. This sequence of holding times defines the branch length of the corresponding tree topology. 

Let the vector $\bfn{}=(n_1,\ldots,n_{m})$ denote the sample sizes at times $\bft^s{}=(t^s_{1},\ldots,t^s_{m})$, with $m$ number of sampling points and $n$ total sample size. The process goes backward in time (from present toward the past): with $t^s_{1}=0$ denoting the present time, and $t^s_{j}>t^s_{j-1}$ for $j=2,\ldots,m$. Let $\bft{}=(t_{n+1}, \ldots, t_{2})$ be the vector of coalescent times with $t_{n+1}=0<t_{n}<...<t_{2}$. 
Note that the subscript in $t_{k}$ is not the current number of extant lineages (often a convention in the coalescent literature)  but the number of lineages that have yet to coalesce. Starting from $t=0$, vectors \bft{} and $\bft^s{}$ partition time into intervals (Figure \ref{fig:het_tree}). An interval ending with a coalescent event, say $t_{k}$, is denoted by $I_{0,k}$; the intervals that end with a sampling time within the interval $(t_{k+1},t_{k})$ are denoted as $I_{i,k}$, where $i\geq 1$ indexes all the sampling events in $(t_{k+1},t_{k})$. Formally, for every $k\in \{2,\ldots,n\}$, we define 
\begin{equation*}
I_{0,k}=[\max\{t_{k+1},t^s_{j}\},t_{k}),\quad \text{ where the maximum is taken over all } t^s_{j}<t_{k},
\end{equation*}
and for every $i\geq 1$ we set
\begin{equation*}
I_{i,k}=[\max\{t_{k+1},t^s_{j-i}\},t^s_{j-i+1}) \text{ with the max taken over all } t^s_{j-i+1}>t_{k+1} \text{ and } t^s_{j}<t_{k}.
\end{equation*}
With $n_{i,k}$ denoting the number of extant lineages during the time interval $I_{i,k}$. Figure~\ref{fig:het_tree} plots an example of a heterochronous genealogy with $\bfn=(1,2,2,2)$, at times $\bft^t=(t^s_1,\ldots,,t^s_4)$ with $t^s_1=0$. In the interval $(t_6, t_5)$ there are two intervals: $I_{1,5}=[t_6,t^s_4)$, $I_{0,7}=[t^s_4,t_5)$.

The vector of coalescent times \bft{} is a random vector whose density with respect to the Lebesgue measure on $\mathbb{R}_+^{n-1}$ depends on two quantities: the coalescent factor $C_{i,k}:=\binom{n_{i,k}}{2}$, and the effective population size $N_e(t)$. The coalescent density can be factorized as the product of the conditional densities of $t_{k-1}$ given $t_k$, i.e. 
\begin{equation}
\label{eq:coal_time}
\hspace{-1.5cm}p(\bft \mid \mathbf{s},\mathbf{n}, N_{e}(t))=\prod_{k=n+1}^3 p(t_{k-1}\mid t_{k},\mathbf{s},\mathbf{n}, N_{e}(t))= \prod_{k=n+1}^3 \frac{C_{0,k-1}}{N_{e}(t_{k-1})} \exp \left\lbrace - \int_{I_{0,k-1}} \frac{C_{0,k-1}}{N_{e}(t)}\di t+\sum^{m}_{i=1}\int_{I_{i,k-1}} \frac{C_{i,k-1}}{N_{e}(t)}\di t\right\rbrace.
\end{equation}
A few remarks. First, the integral over $I_{i,k-1}$ accounts for the probability of no coalescence during $I_{i,k-1}$. It is zero if there are less than $i$ sampling times between $t_k$ and $t_{k-1}$. Second, conditionally on \bfs{}, \bfn{} and $t_k$, the coalescent factors can be computed exactly and $N_e(t)$ is the only unknown parameter, sampling times are assumed fixed.

Coalescent times can be alternatively viewed as the realization of
an inhomogeneous point process with rate $C(t) N_e(t)^{-1}$, with the coalescent factor $C(t)$ being defined for all $t \geq 0$ by the notation above. This alternative view allows us to frame the problem of inferring $N_e(t)$ as that of inferring the intensity function of an inhomogeneous point process. \cite{pal13} is an example of how this representation is useful in inference and simulations.


\subsection{Some priors for the effective population size}
\label{sec:priors}

Markov random field-based priors on the log effective population trajectory allows smoothing without careful modeling of change points. They are computationally tractable thanks to the sparsity assumption in the covariance matrix of the field \citep{rue2005gaussian}. All MRF-based priors for phylodynamic inference share the assumption that the trajectory $N_e(t)$ is an unknown continuous function. The integral in \eqref{eq:coal_time} is numerically approximated by the Riemann sum at a regular grid of $M+1$ points $(k_i)_{1:M+1}$, and one assumes that the trajectory $N_e(t)$ is well approximated by $\sum^{M+1}_{i=1}\exp \theta_i 1(t \in (k_{i},k_{i+1}))$, with $\boldsymbol{\theta}=(\theta_i)_{1:M}$. We stress that neither the grid cell boundaries $(k_i)_{1:M+1}$ nor $M$ depend on \bft{} and $N_e(t)$, with the choice of $M$ commonly based on $n$ \citep{fau20}. A description of the discretized coalescent log-likelihood $\mathcal{L}(\boldsymbol{\theta}|\bft)$ is given in detail in \cite{pal12} and \cite{fau20}.

The Horseshoe Markov random field prior (HSMRF) for $\boldsymbol{\theta}$ \citep{fau20} assumes that the $p$th order forward differences of $\boldsymbol{\theta}$ are independent and Horseshoe distributed \citep{car10}, i.e.
\begin{equation}
\label{eq:hsmrf}
\Delta^p \theta_i | \tau_i \sim N(0,\tau^2_i) \,\ \,\ \,\ \,\ \tau_i | \gamma \sim C^+(0,\gamma) \,\ \,\ \,\ \,\ \gamma|\zeta \sim C^+(0,\zeta) \,\ \,\ \,\ \,\ \text{for} \,\, p+1\leq i \leq M-1,
\end{equation}
where $C^+(0,a)$ is the standard half-Cauchy distribution with positive support with scale parameter $a$, $\tau_i$ are the local shrinkage parameters and $\gamma$ is the global smoothing parameter. To completely specify the prior, one sets $\theta_1 \sim N(\mu,\sigma_1^2)$, and for $p\geq 2$, the first p values of the field have running order q difference priors as follows:
$$\Delta^q \theta_q | a_q \tau_q \sim N(0, a_q\tau^2_q) \,\ \,\ \,\ \,\ \tau_q | \gamma \sim C^+(0,\gamma) \,\ \,\ \,\ \,\ \gamma|\zeta \sim C^+(0,\zeta) \,\ \,\ \,\ \,\ \text{for} \,\, 1\leq q \leq p-1,$$
with $a_q=2^{-(p-q)/2}$. 
As it is common in the trend filtering literature \citep{kim09}, only orders $1$ and $2$ are typically employed in applications. 

A related prior consists in assuming that the $p$th order forward difference of $\boldsymbol{\theta}$, more precisely the vector $(\theta_1, \Delta^1\theta_1,\ldots,\Delta^p\theta_p,\Delta^p\theta_{M-1})$ is distributed as a GMRF with mean vector $\boldsymbol{\mu}$ and covariance matrix $\boldsymbol{Q}(\gamma)$ corresponding to: 
\begin{equation}
\label{eq:gmrf}
\Delta^p \theta_i | \gamma \sim N(0,\gamma^2) \,\ \,\ \,\ \,\  \gamma|\zeta \sim C^+(0,\zeta) \,\ \,\ \,\ \,\ \text{for} \,\, p+1\leq i \leq M-1,
\end{equation}
$\theta_1 \sim N(\mu,\sigma_1^2)$, and for $1\leq q \leq p-1$  we set $\Delta^q \theta_q | a_q \gamma \sim N(0, a_q\gamma^2)$. A common alternative to the half-Cauchy distribution on $\gamma$ is a Gamma prior, e.g. in \cite{pal12}. We will employ both formulations in our implementations.

A fully nonparametric prior on $\log N_e(t)$ has been studied by \cite{pal13}, who proposed a Gaussian process prior on the log effective population size. The advantage of this approach is that no grid needs to be specified a priori. In applications, we believe that the GMRF, the discretized version of this prior, achieves a comparable empirical performance. 


\subsection{Preferential sampling}

Preferential sampling arises when the process that determines the locations of the data (i.e. sampling process) and the process under study are stochastically dependent. The notion was introduced by \cite{dig10} who show that not accounting for this effect leads to biased inference as a result of the model misspecification. On the other hand, a correctly specified sampling model can lead to more accurate estimates. 

In phylodynamics, preferential sampling arises when the sampling process depends on $N_e(t)$. \cite{vol14} provide the first evidence that coalescent-based inference under a misspecified sampling process can be biased. They propose a new estimator tailored to a coalescent process with exponentially growing effective population size and a sampling process with rate linearly dependent on $N_e(t)$. They show that the estimator obtained by correctly modeling the sampling process is more accurate than the standard coalescent estimator.

\cite{kar16} assume that $N_e(t)$ is a continuous function and the sampling process is a Poisson process with rate $\lambda(t)=\exp(\beta_0)N_e(t)^{\beta_1}$, for $\beta_0,\beta_1\geq0$, i.e. the rate $\lambda(t)$ is proportional to the effective population size. This model is parsimonious, capturing a variety of scenarios with two parameters: with $\beta_1=1$, the rate is a constant times $N_e(t)$, on the opposite side of the spectrum, with $\beta_1=0$, one models uniform sampling. Another advantage is that little assumptions are made on $N_e(t)$. However, the parametric assumptions on $\lambda(t)$ make the sampling dates strongly informative about $N_e(t)$. This situation can be problematic in the case of sampling dates errors. Moreover, under no preferential sampling or under a different rate $\lambda(t)$ (model misspecified), it constitutes a relevant model misspecification, and leads to estimation biases; see for example Figure \ref{fig:traj/est} in Section \ref{sec:sim}. 
\cite{kar20b} address some of the limitations of the parametric model by including time-varying covariates into the Poisson process rate: $\lambda(t)=\exp(\beta_0)N_e(t)^{\beta_1} +\boldsymbol{\beta}'\textbf{X}(t)$, where $\textbf{X}$ is a vector of covariates and $\boldsymbol{\beta}'$ the corresponding linear coefficients. Here a covariate can be for example a dummy variable indicating a change in sampling protocols, or when a new sampling center joined the study. The term $\boldsymbol{\beta}'\textbf{X}$ adds more flexibility to the parametric dependence enforced by $\exp(\beta_0)N_e(t)^{\beta_1}$. Clearly, this extension implies the availability of covariates informative on the sampling design. 

\cite{par20} introduce the 
epoch sampling skyline plot (ESP) estimator that allows for a more flexible dependence of the sampling process on the effective population size. More specifically, the ESP method assumes that $N_e(t)$ is a piecewise-constant function with $r$ segments described by the vector $(N_1,\ldots, N_{r})$ of r parameters, and time is further partitioned in $d$ epochs, such that in epoch $i$ and segment $r$ the sampling process is a Poisson process with rate $\beta_{i} N_{j}$, where $(\beta_1,\ldots, \beta_d)$ is a vector of $d$ parameters. The vector $(\beta_1,\ldots, \beta_d)$ modulates the dependence of the sampling process on the effective population size, assuming that the dependence changes across $d$ epochs.  This is a notable advantage over the parametric model of \cite{kar16}: one can model a variety of realistic time-varying sampling protocols, or simply deal with sampling discontinuities typical of outbreaks. We conjecture that higher flexibility in the preferential model reduces the risk of model misspecification and bias when the preferential sampling assumption does not hold. 

In the ESP, the endpoints of the $r$ segments coincide with a subset of the coalescent times \bft{}. Similarly, the boundary points of the $d$ epochs are determined by a subset of the sampling times.  The number of segments $r$ and epochs $d$, as well as their lengths, need to be determined or inferred. These choices affect the ESP estimates heavily. The authors implement a frequentist and a Bayesian version with independence assumptions in $(N_1,\ldots, N_{p})$ and $(\beta_1,\ldots, \beta_d)$, leading to estimates with high variance, a characteristic feature of skyline plot-type estimators.

\section{Adaptive preferential sampling}


In the adaptive preferential sampling framework, the sampling times $(s_i)_{2:n}$ are determined by the jumps of an inhomogenous Poisson process with rate $\lambda(t)=\beta(t) N_e(t)$, with $N_e(t)$ effective population size, and $\beta(t)$, a function modulating the linear dependence between $\lambda(t)$ and $N_e(t)$, and $s_1$ is fixed to $0$.  We assume that both $\beta(t)$ and $N_e(t)$ are unknown continuous functions. To numerically approximate the integrals in \eqref{eq:coal_time}, we resort to the approximation sketched in  in Section \ref{sec:priors} and detailed in \cite{pal12}. We employ  the regular grid $(k_i)_{1:M+1}$ and assume that $N_e(t)$ is governed by parameters $\boldsymbol{\theta}=(\theta_i)_{i=1:M}$. Similarly, to model the time-varying rate $\lambda(t)$, we assume that $\beta(t)$ is governed by parameters $\boldsymbol{\alpha}=(\alpha_i)_{i=1:M'}$, where
$M' = \min_{i} \{k_{i+1} : k_{i+1}>s_n\}$ and $\beta(t)\approx \exp \alpha_i$ for $t \in (k_i, k_{i+1}]$. 

A few preliminary remarks. Modeling the $\log \beta(t)$ ensures that $\beta(t)\geq 0$. $M'$ is not related to the number of epochs $d$ in ESP: it is solely determined by $s_n$ and the grid $(k_i)_{1:M+1}$, which in turn does not depend on \bft{}. We have by definition $M'<M$ to ensure that $\beta(t)$ is not modeled after the last sampling time. Lastly, after discretizing $\beta(t)$, we can write the log-likelihood contribution of the sampling process as
\begin{equation}
\label{eq:samp_proc}
\mathcal{L} (\boldsymbol{\alpha},\boldsymbol{\theta} | \bfs) = \sum_{i=1}^{M'} \Big[\Big|\{s_i:s_i \in (k_i,k_{i+1}]\}\Big|(\alpha_i + \theta_i +\log \Delta_i ) - \exp\{\alpha_i \, \theta_i\} \Delta_i\Big],
\end{equation}
where $\Delta_i=k_{i+1}-k_i$ and the first interval $[k_1,k_2]$ is closed to include $s_1$. Through the term $|\{s_i:s_i \in (k_i,k_{i+1}]\}|$, the discretized log-likelihood \eqref{eq:samp_proc}  allows to naturally account for multiple sampling times collected at once, reconciling this model with the description of the heterochronous coalescent given in Section \ref{sec:coal}, in particular the definition of vectors $\bft^s$ and \bfn. 

Here, we model both $\boldsymbol{\alpha}$ and $\boldsymbol{\theta}$ through Markov random field priors, either HSMRFs or GMRFs. This allows us to make minimal assumptions on $N_e(t)$  and $\beta(t)$: the choice of the grid practically depends solely on the sample size and no major assumptions are made on the underlying sampling process. The choice of prior for $N_e(t)$ follows from the well-studied characteristics of the two priors discussed in Section \ref{sec:priors}. Under the HSMRF prior on $\beta(t)$, one can model situations in which there are sharp changes in the sampling design (both first and second orders). Under the GMRF prior on $\beta(t)$, one favors smooth sampling designs, a situation which is also desirable when one does not have exact knowledge of the underlying sampling protocol. Note that the choice of field and order of the priors can be disjoint: for example, one can place a HSMRF of order $1$ prior on $N_e(t)$ and a GMRF of order $2$ prior on $\beta(t)$.

To formalize, Bayesian phylodynamic inference under adaptive preferential sampling can be written in the most general form as 
\begin{align}
\label{eq:model}
\bft{}|\boldsymbol{\theta} , \bfn, \bfs  \sim \text{Coalescent model}   \,\ \,\ &\,\ \,\ \,\ \bfs | \boldsymbol{\theta}, \boldsymbol{\alpha}, n \sim \text{Poisson process}\\
\boldsymbol{\theta}|\boldsymbol{\tau},\gamma \sim \text{HSMRF-}p_1  \,\ \text{or}  \,\ \,\ \boldsymbol{\theta}| \gamma \sim \text{GMRF-}p_1 \,\ \,\ \,\ &\,\ \,\ \,\ \,\ \boldsymbol{\alpha}|\boldsymbol{\psi},\xi \sim \text{HSMRF-}p_2 \,\  \text{or}  \,\ \,\ \boldsymbol{\alpha}| \xi \sim \text{GMRF-}p_2, \nonumber
\end{align}
where $\xi$ is the global smoothing parameter of the MRF on $\boldsymbol{\alpha}$, $\boldsymbol{\psi}$ is the vector of local shrinkage parameter of the HSMRF prior on $\boldsymbol{\alpha}$, $p_1$ and $p_2$ are the orders of the respective MRFs. We will refer to any combination of priors above as the adaptive preferential model.

Note that the adaptive preferential model differs notably from the framework of the ESP estimator by the fact that the parameter vectors $\boldsymbol{\theta}$ and $\boldsymbol{\alpha}$ are each dependent, the grid at which they are defined does not depend on $\bft{}$, and these priors favor smooth estimates. 


\subsection{Inference}

\textit{Posterior distributions.} Under the assumption that \bft{} and \bfs{} are known, and that we place HSMRF priors on both $\boldsymbol{\alpha}$ and $\boldsymbol{\theta}$, the posterior distribution of model parameters could be readily computed
$$\pi(\boldsymbol{\alpha}, \boldsymbol{\theta}, \boldsymbol{\psi}, \boldsymbol{\tau},\gamma, \xi | \bft{},\bfs) \propto \mathcal{L} (\boldsymbol{\alpha},\boldsymbol{\theta} | \bfs) \mathcal{L} (\boldsymbol{\theta} | \bft) \pi(\boldsymbol{\theta}|\boldsymbol{\tau}) \pi(\boldsymbol{\tau}|\gamma) \pi(\gamma|\zeta_1) \pi(\boldsymbol{\alpha}|\boldsymbol{\psi}) \pi(\boldsymbol{\psi}|\xi) \pi(\xi|\zeta_2),$$
where $ \mathcal{L} (\boldsymbol{\theta} | \bft)$ is the discretized coalescent log-likelihood. Under GMRF priors on $\boldsymbol{\alpha}$ and $\boldsymbol{\theta}$, the posterior would be 
$$\pi(\boldsymbol{\alpha}, \boldsymbol{\theta}, \gamma, \xi | \bft{},\bfs) \propto \mathcal{L} (\boldsymbol{\alpha},\boldsymbol{\theta} | \bfs) \mathcal{L} (\boldsymbol{\theta} | \bft) \pi(\boldsymbol{\theta}|\gamma)  \pi(\gamma|\zeta_1) \pi(\boldsymbol{\alpha}|\xi) \pi(\xi|\zeta_2).$$
For our analysis, we fixed the pair $(g,\bft)$, which can be estimated by other methods such as the Maximum clade credibility tree of the posterior distribution of the genealogy. In order to approximate the posterior distribution we use two methods: Hamiltonian MCMC and Integrated Nested Laplace approximation (INLA, \cite{rue09}), in particular for Hamiltonian MCMC, we rely on \texttt{Stan} \citep{car17}. The hyperparameters $\zeta_1$ and $\zeta_2$ are as described in Appendix B of \cite{fau20} (their method is suitably adapted to $\zeta_2$, the global smoothing parameter of the MRF on $\boldsymbol{\alpha}$).

\noindent \textit{INLA approximation.}  Posterior inference from latent Gaussian models can be achieved by approximating posterior marginal distributions via Laplace approximations. INLA allows us to replace MCMC entirely and approximate the posterior marginals of model parameters when our model is based on GMRF priors. What follows is largely based on \cite{pal12}, who discuss INLA for GMRFs in phylodynamics. We extend it here to include the adaptive preferential sampling priors.

INLA approximates posterior marginals $\pi(\gamma, \xi |\bft{},\bfs{}), \pi(\theta_i|\bft{},\bfs{})$ for $1\leq i\leq M$, and  $\pi(\alpha_j|\bft{},\bfs{})$ for $1\leq j\leq M'$. The posterior marginal distribution of hyperparameters is 
$$\widehat{\pi}(\gamma, \xi |\bft{},\bfs{}) \propto \frac{\pi(\gamma,\xi ,\boldsymbol{\theta},\boldsymbol{\alpha},\bft{}.\bfs{})}{\widehat{\pi_G}(\boldsymbol{\theta},\boldsymbol{\alpha} | \gamma,\xi,\bft{},\bfs{})}\Bigg|_{\boldsymbol{\alpha}=\boldsymbol{\alpha}^*(\xi,\gamma),\boldsymbol{\theta}=\boldsymbol{\theta}^*(\xi,\gamma)},$$
where  $\widehat{\pi_G}(\boldsymbol{\theta},\boldsymbol{\alpha} | \gamma,\xi,\bft{},\bfs{})$ is the Gaussian approximation of $\pi(\boldsymbol{\theta},\boldsymbol{\alpha} | \gamma,\xi,\bft{},\bfs{})$ obtained from a Taylor expansion around its modes $\boldsymbol{\theta}^*(\xi,\gamma)$ and $\boldsymbol{\alpha}^*(\xi,\gamma)$ (modes can be computed through any optimization algorithm, e.g. Newton-Raphson).

The approximation of the marginal distributions of the MRFs $\pi(\theta_i|\bft{},\bfs{})$ and $\pi(\alpha_j|\bft{},\bfs{})$ are
$$\hspace{-1cm} \widehat{\pi}(\theta_i |\gamma,\xi,\bft{},\bfs{}) \propto \frac{\pi(\gamma,\xi ,\boldsymbol{\theta},\boldsymbol{\alpha},\bft{}.\bfs{})}{\widehat{\pi_{GG}}(\boldsymbol{\theta_{-i}},\boldsymbol{\alpha} | \gamma,\xi,\bft{},\bfs{})}\Bigg|_{\boldsymbol{\alpha}=\boldsymbol{\alpha}^*,\boldsymbol{\theta_{-i}}=\boldsymbol{\theta_{-i}}^*}   \text{and}  \,\  \,\ \,\,   \widehat{\pi}(\alpha_i |\gamma,\xi,\bft{},\bfs{}) \propto \frac{\pi(\gamma,\xi ,\boldsymbol{\theta},\boldsymbol{\alpha},\bft{}.\bfs{})}{\widehat{\pi_{GG}}(\boldsymbol{\theta},\boldsymbol{\alpha_{-i}} | \gamma,\xi,\bft{},\bfs{})}\Bigg|_{\boldsymbol{\alpha_{-i}=\boldsymbol{\alpha_{-i}^*,\boldsymbol{\theta}}=\boldsymbol{\theta}}^*},$$
where $\widehat{\pi_{GG}}(\boldsymbol{\theta},\boldsymbol{\alpha_{-i}} | \gamma,\xi,\bft{},\bfs{})$ is the Gaussian approximation of $\pi(\boldsymbol{\theta},\boldsymbol{\alpha_{-i}} | \gamma,\xi,\bft{},\bfs{})$ obtained from a Taylor expansion at $(\boldsymbol{\alpha_{-i}},\boldsymbol{\theta})=\E_G[\boldsymbol{\theta},\boldsymbol{\alpha_{-i}} | \gamma,\xi,\bft{},\bfs{}]$, where $E_G$ denotes the expected value w.r.t. $\widehat{\pi_{G}}(\boldsymbol{\theta},\boldsymbol{\alpha} | \gamma,\xi,\bft{},\bfs{})$. Analogously, we can define the same term for $\widehat{\pi_{GG}}(\boldsymbol{\theta_{-i}},\boldsymbol{\alpha} | \gamma,\xi,\bft{},\bfs{})$. Given $\widehat{\pi}(\theta_i |\gamma,\xi,\bft{},\bfs{})$ and $\widehat{\pi}(\alpha_i |\gamma,\xi,\bft{},\bfs{})$, we use $\widehat{\pi}(\gamma, \xi |\bft{},\bfs{})$ to integrate out $\gamma$ and $\xi$ and obtained the desired  distributions.

\section{Simulations}
\label{sec:sim}

We rely on simulations to evaluate the performance of the adaptive preferential sampling (adaPref) method in estimating the effective population size trajectory in the presence of ``strong" preferential sampling and under ``weak" preferential sampling in which sampling is preferential only during some time periods. 
We then evaluate the sensitivity of adaPref posterior inference to the choice of the MRF (GMRF vs HSMRF) priors and their orders ($1$ vs $2$). We compare the performance of adaPref to alternative methods with and without preferential sampling. In the supplementary material we study the approximation error incurred using INLA  in place of MCMC. Also, although $\beta(t)$ is a parameter of no direct scientific interest, we 
deem to successfully recover all model parameters, and we study how well our model infer $\beta(t)$.

\noindent {\it Simulation setup.} For each simulated dataset, we estimated adaPref posteriors using 16 different models with all possible combinations of GMRFs and HSMRFs of orders 1 and 2 per parameter $N_{e}(t)$ and $\beta(t)$. We compare these models to those obtained by ignoring preferential sampling implemented in the the \texttt{R} packages \texttt{spmrf} \citep{fau20} and \texttt{phylodyn} \citep{phylodyn}. 

We use \texttt{smprf} to estimate the posterior of $N_e(t)$, without preferential sampling, with GMRF and HSMRF priors of orders $1$ and $2$. Each posterior distribution is approximated with $2000$ MCMC samples, obtained running $4$ chains, each with $2000$ iterations, with a burnin of $1000$ iterations and thinning every $2$ iterations. Both HMC implementations (adaPref priors and \texttt{spmrf}) rely on \texttt{Stan}, in particular the \texttt{R} interface \texttt{rstan} \citep{stan18}. For our implementations, we use the same settings used by \cite{fau20}.

In addition, we use INLA approximations for models that employ GMRF priors. We used \texttt{R-INLA} \citep{rue09} for our implementation of adaPref with GMRF order 1 priors (GMRF1) on both $N_e(t)$ and $\beta(t)$. We compare GMRF1 adaPref with the  GMRF1 prior with preferential sampling of \cite{kar16} (parPref), and the the  GMRF1 prior without preferential sampling (noPref) \citep{pal12}.

For each dataset, we test the performance of all models through a set of commonly used summary statistics. For a regular grid of time points $(v_i)_{i:K}$, we consider the sum of relative errors: $SRE=\sum^{K}_{i=1}\frac{|\widehat{N}_e(v_{i})-{N_e(v_{i})}|}{{N_e(v_{i})}}$, where  $\widehat{N}_e(v_{i})$ is the posterior median of $N_e$  at time $v_{i}$; the mean relative width:
$MRW=\frac 1 K \sum^{K}_{i=1}\frac{|\hat{N}_{97.5}(v_{i})-\hat{N}_{2.5}(v_{i})|}{N(v_{i})}$,
where $\hat{N}_{97.5}(v_{i})$ and $\hat{N}_{2.5}(v_{i})$ are respectively the $97.5\%$  and $2.5\%$ quantiles of the posterior distribution of $N(v_{i})$; lastly,  the envelope measure 
$ENV= \frac 1 K \sum^{K}_{i=1}\mathbf{1}_{\{\hat{N}_{2.5}(v_{i})\leq N_e(v_{i}) \leq  \hat{N}_{97.5}(v_{i})\}},$
which measures the proportion of the curve that is covered by the $95$\% credible region. We fix $K=100$, $v_1=0$ and $v_k=.8\, t_2$. We compute the Watanabe Aikake information criteria (WAIC,\cite{wat10}) implemented in the \texttt{R} package \texttt{loo} \citep{veh17}, for model comparison across the $20$ models computed through \texttt{rstan}.

\begin{figure}[!t]
	\centering
	\includegraphics[scale=0.58]{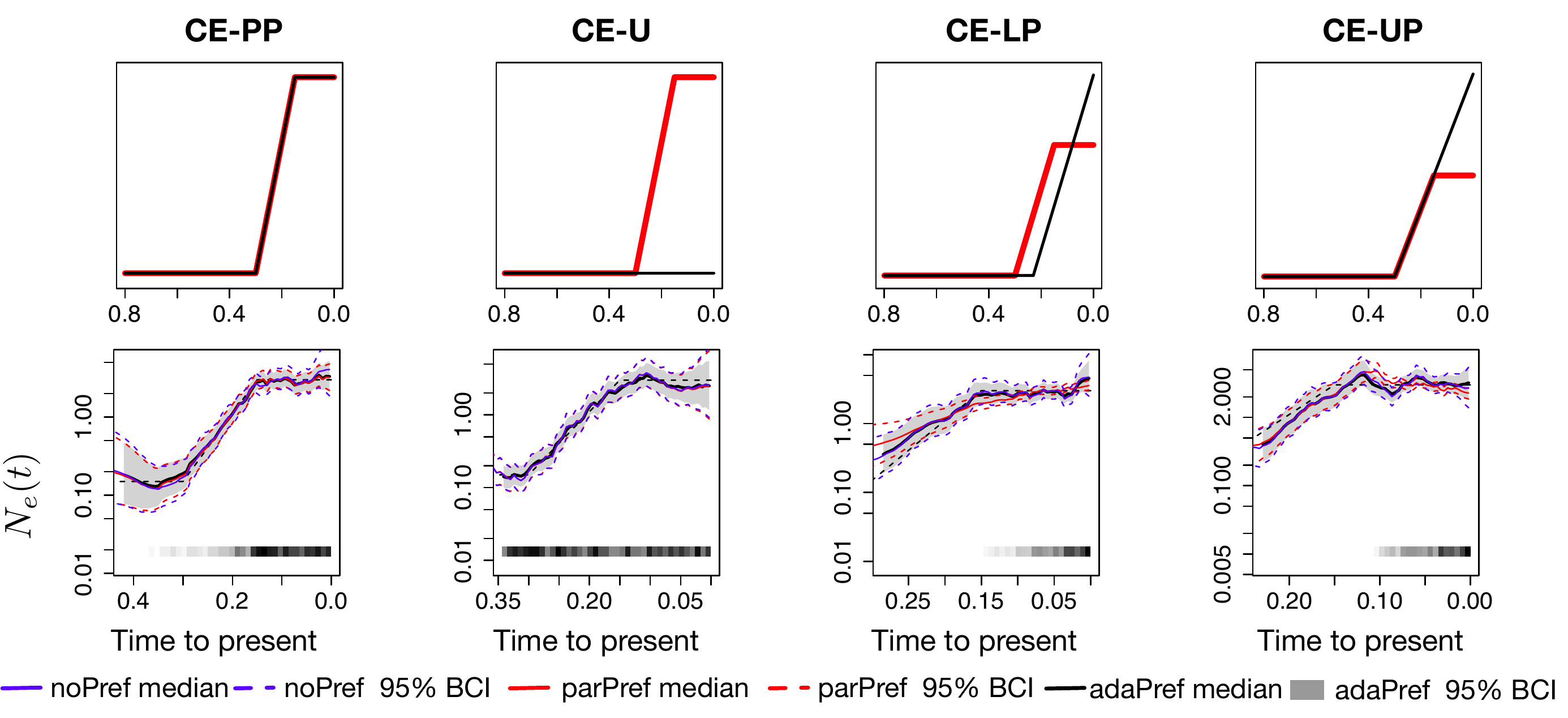}
	\caption{\small{\textbf{Simulations from constant-exponential trajectories and posterior inferences of $N_e(t)$.} First row panels depict the simulated log-effective population size (red) and log-sampling intensity (black) trajectories (up to a constant). Second row panels depict posterior estimates for the four simulation scenarios from a single simulated genealogy picked at random with $n=500$ tips. All models used GMRF of order 1 priors and posterior inference is approximated with INLA. The posterior medians of adaPref are depicted as solid black curves and the 95\% Bayesian credible regions are depicted as shaded areas. Posterior medians of parPref and noPref are depicted respectively as solid blue and red curves, and the 95\% Bayesian credible regions are depicted by the corresponding dashed curves. \bfn{} and \bfs{} are depicted by the heat maps at the bottom of the last four panels: the squares along the time axis depict the sampling times, while the intensity of the black color depicts the number of samples. The true trajectories are depicted as a black dashed curves.}}
	\label{fig:traj/est_exp}
\end{figure}

\begin{figure}[!t]
	\centering
	\includegraphics[scale=0.58]{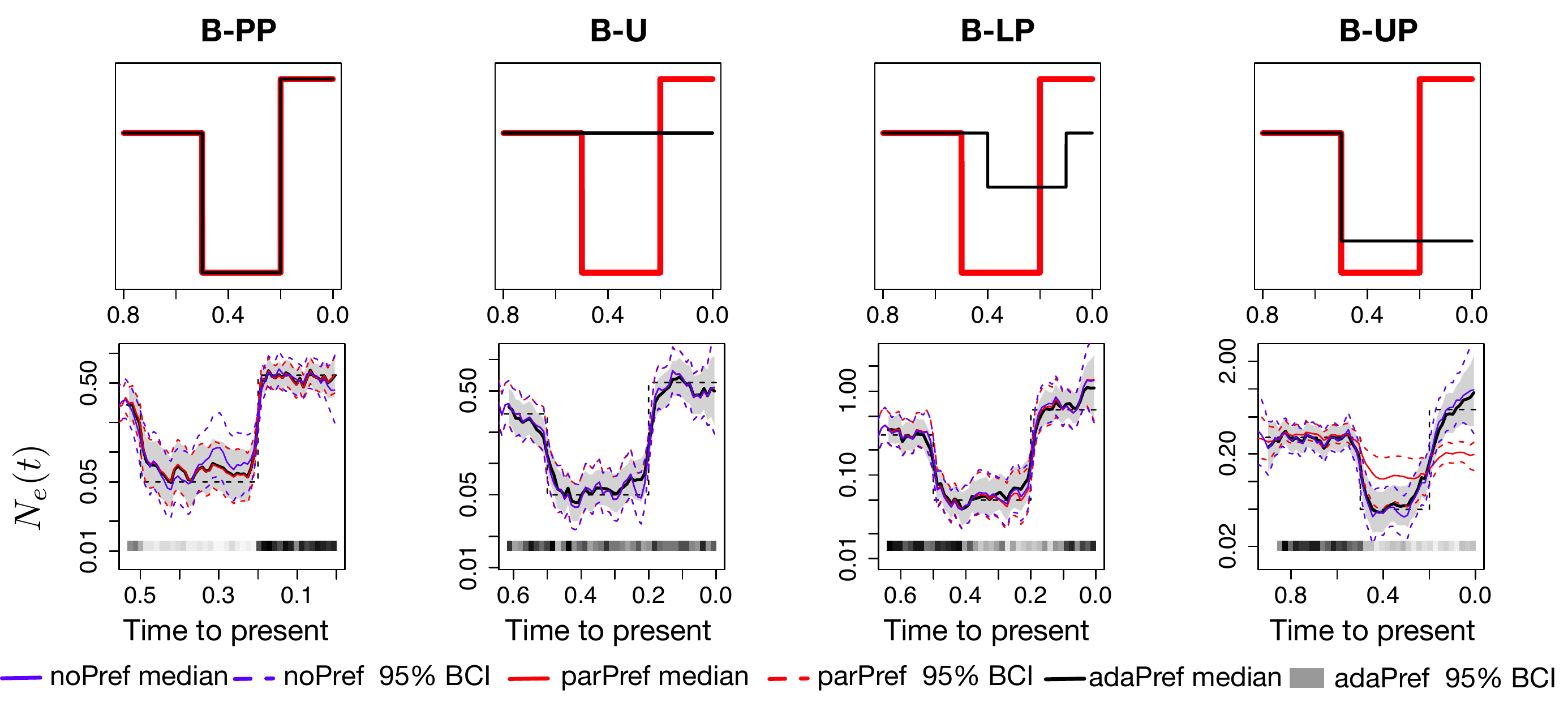}
	\caption{\small{\textbf{Simulations from bottleneck trajectories and posterior inferences of $N_e(t)$.} First row panels depict the simulated log-effective population size (red) and log-sampling intensity (black) trajectories (up to a constant). Second row panels depict posterior estimates for the four simulation scenarios and from a single simulated genealogy picked at random with $n=500$ tips. All models used GMRF of order 1 priors and posterior inference is approximated with INLA. The posterior medians of adaPref are depicted as solid black curves and the 95\% Bayesian credible regions are depicted by shaded areas. Posterior medians of parPref and noPref are depicted respectively as solid blue and red curves, and the 95\% Bayesian credible regions are depicted by the corresponding dashed curves. \bfn{} and \bfs{} are depicted by the heat maps at the bottom of the last four panels: the squares along the time axis depict the sampling times, while the intensity of the black color depicts the number of samples. The true trajectories are depicted as a black dashed curves.}}
	\label{fig:traj/est}
\end{figure}

\noindent {\it Data. } We simulate genealogies under two population size trajectories: a piece-wise constant and exponential trajectory (CE), and a bottleneck trajectory (B). For the sampling protocols, we simulate sampling trajectories that resemble situations encountered in applications:  a sampling protocol proportional to $N_e(t)$ (PP), uniform (U),  a ``lagged" response to changes in $N_e(t)$ (LP), and a situation where only some segments of the sampling trajectory are preferential (UP). The combination of the two acronyms will be used in the plots, \textit{e.g} B-U refers to bottleneck trajectory and uniform sampling. The first rows of Figures \ref{fig:traj/est_exp}-\ref{fig:traj/est} depict the $N_{e}$ (red) and $\lambda(t)$ (black) trajectories (up to a scaling constant and in log-scale) of the eight simulation scenarios considered. Exact specifics of the trajectories used are given in the supplementary material. 

\begin{figure}[!t]
	\centering
	\includegraphics[scale=0.6]{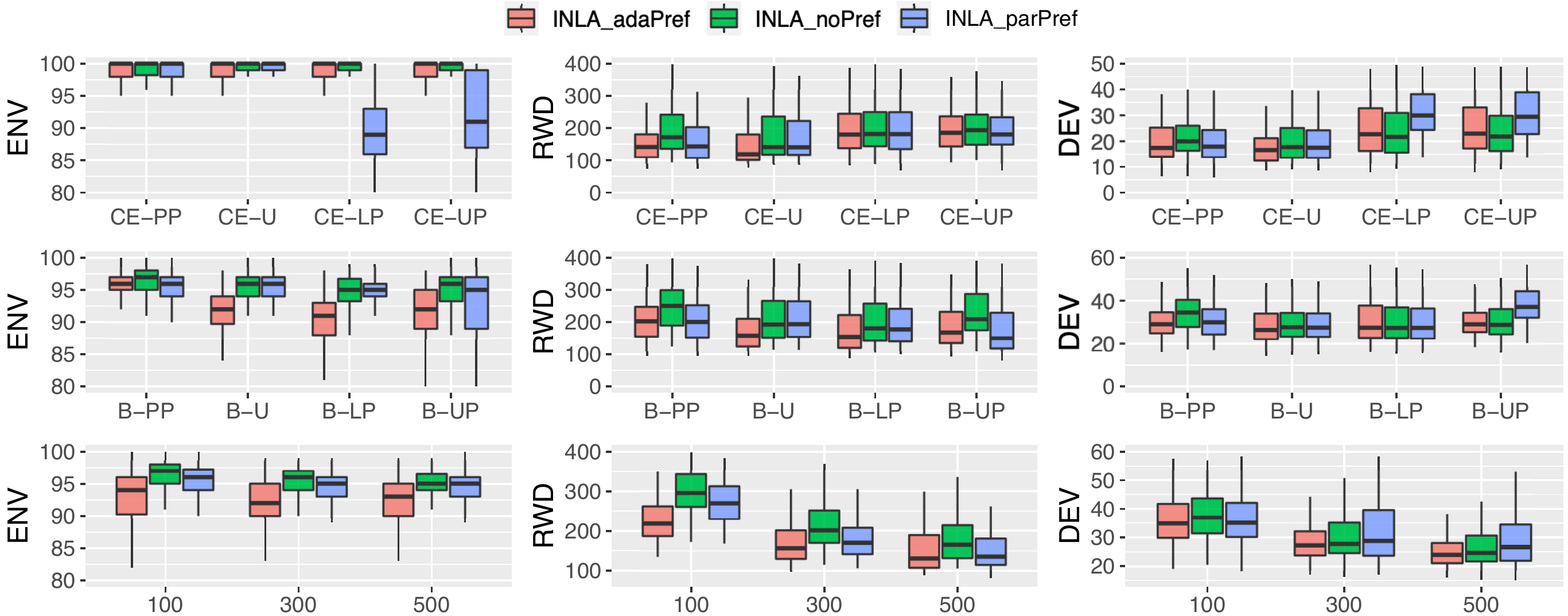}
	\caption{\small{\textbf{Summary statistics of  parPref, noPref and adaPref inference of $N_{e}(t)$ approximated with INLA.} In the first two rows, each box refers to  one method (color in the legend) and depicts the distribution of the $150$ estimated statistics for each simulation trajectory ($50$ datasets for each sample size, $150$ in total) based on $N_e(t)$ posterior: ENV, first column; RWD, second column; DEV, third column. In the third row, the grouping is done according to the sample size: each box is based on $400$ simulated genealogies ($50$ genealogies for each of the eight trajectories). In the legend, INLA\_adaPref is our method, INLA\_noPref is the method in \cite{pal12}, INLA\_parPref is the method of \cite{kar16}. All models used GMRF of order 1 priors.}}
	\label{fig:boxINLA}
\end{figure}

\begin{figure}[!t]
	\centering
	\includegraphics[scale=0.58]{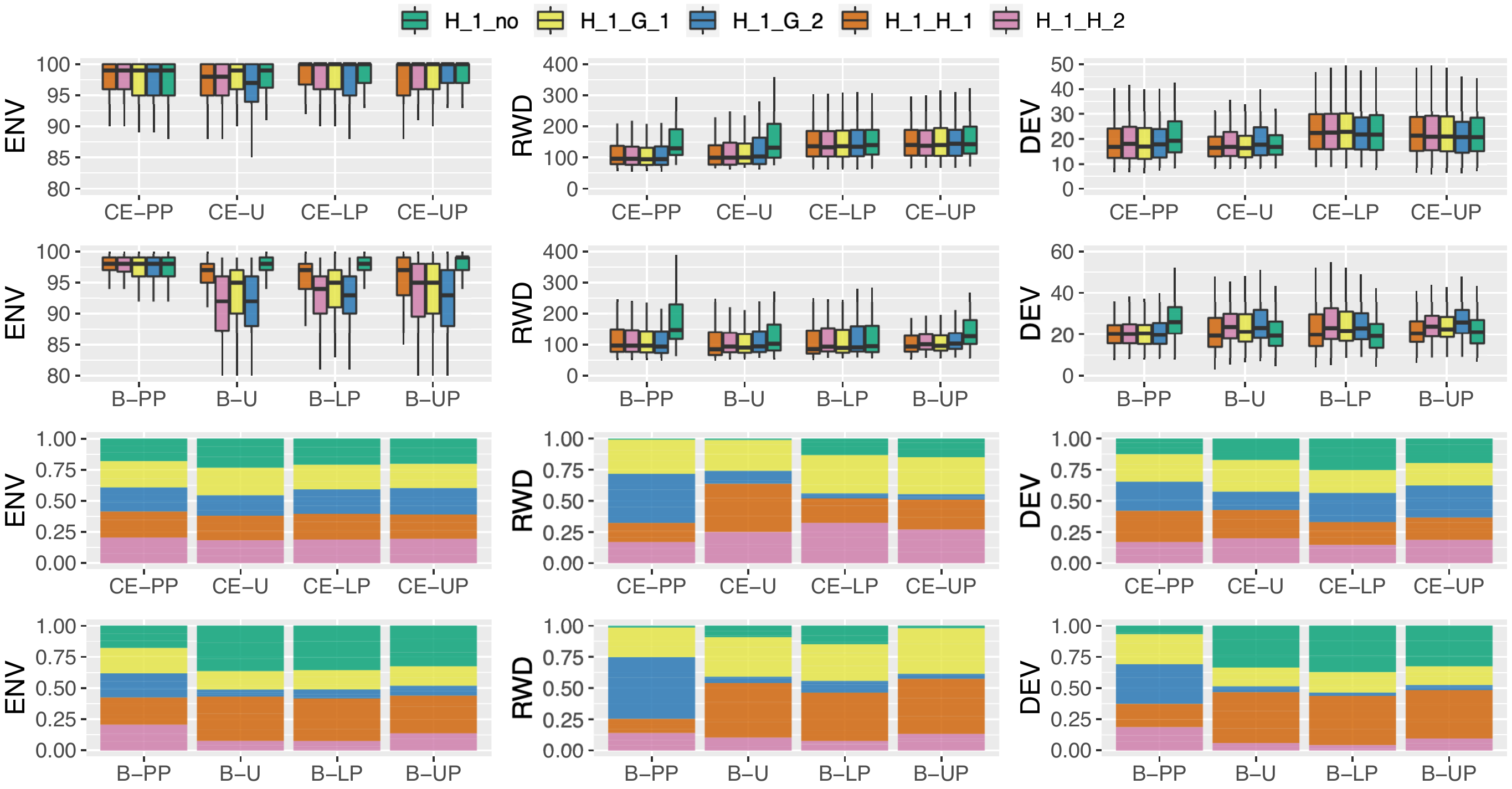}
	\caption{\small{\textbf{Summary statistics of noPref and adaPref inference of $N_e(t)$ on simulations with different MRF priors.} In the boxplots (top two rows), each box refers to one method (color in the legend) and depicts the distribution of the $150$ estimated statistics  for each simulation trajectory ($50$ datasets for each sample size, $150$ in total) based on $N_e(t)$ posterior: ENV, first column; RWD, second column; DEV, third column.  In the stacked bar plots (bottom two rows), each bar refers to a simulation scenario and each sub-bar refers to a model (color legend). The sub-bar width represents the percentage of simulated datasets for which the model under consideration is in the top two best performance for the corresponding summary statistics ( highest ENV, lowest RWD, and lowest DEV). In the legend, the first capital letter refers to the type of prior on $N_e(t)$, the second one on $\beta(t)$, with H denoting the HSMR prior, G the GMRF prior; the two numbers refer to the corresponding orders. }}
	\label{fig:stack+box}
\end{figure}


We simulate both the coalescent times and the sampling times from their corresponding inhomogeneous Poisson processes using the Lewis-Shedler thinning algorithm \citep{lew79,pal13}. We consider three sample sizes $n$ in ($100,300,500$) and $50$ simulations  for each combination of trajectories and sample size. We estimated posteriors with the twenty-three models for each simulated genealogy. The code for reproducing the simulation study is available at \texttt{https://github.com/lorenzocapp/adapref} as a \texttt{R} package.

%

\noindent {\it Results.}  We first analyze a single simulated genealogy with $n=500$ tips from each simulation scenario and approximate posterior marginals with INLA assuming GMRFs of order 1 priors.
The four panels of Figure \ref{fig:traj/est_exp} second row 
depict the posterior medians and $95$\% BCI of $N_e(t)$ for the constant-exponential trajectory (CE). 
Our adaPref results are depicted in black and grey scale, the parPref method in red, and the noPref in blue. 
Each column corresponds to \lc{each}{a} sampling protocol. All posterior medians are very similar and close to the truth (black dashed line) except for parPref (red) in the last two scenarios. Indeed, the last two scenarios correspond to the cases in which the preferential sampling assumption of parPref is violated. ParPref and adaPref show similar credible region widths in the case of proportional preferential sampling (first column), however, adaPref consistently shows narrower credible regions across sampling protocols.
Figure \ref{fig:traj/est} shows the same type of comparisons for the bottleneck trajectory (B). The posterior median and credible intervals obtained with parPref are particularly off during the periods of no preferential sampling in the last simulation scenario (fourth column). In all other sampling scenarios, all methods have very similar posterior medians, however again, adaPref shows narrower credible regions while keeping high coverage across sampling protocols. 


We now discuss the accuracy of the $N_{e}(t)$ estimators obtained with the three different models: adaPref, noPref, and parPref (approximated with INLA and using GMRF-1 priors) in the four sampling protocols for CE and B trajectories but now summarizing the estimations obtained from all 150 simulated genealogies.

Figure \ref{fig:boxINLA} top two rows plot the ENV, RWD, and DEV summary statistics obtained from the CE (first row) and the B (second row) trajectories. The adaPref model (red boxplots) has the best mean performance in six out of the eight scenarios in terms of both RWD and DEV (B-PP, B-U, all CE scenarios). In the last two scenarios (B-LP, B-UP), the parPref model achieved the lowest mean RWD and noPref the lowest mean DEV. Surprisingly, the adaPref model outperforms parPref also in the CE-PP and B-PP scenarios, where the parametric assumptions are met.

The adaPref model is more heavily parametrized and one may be lead to think that the performance of the adaPref estimator is affected by the sample size. Figure \ref{fig:boxINLA} last row panels depict the boxplots of the three statistics considered, now grouping simulations by sample size. There is no detectable sample size effect: the relative performance of the estimators is roughly similar as $n$ increases. The adaPref estimator is the best performing according to DEV and RWD averaging over all the simulation scenarios jointly.

We now assess the sensitivity of different MRF priors on $\beta(t)$ and $N_{e}(t)$ parameters in the adaPref model and compare the $N_{e}(t)$ adaPref estimators to the noPref estimators according to ENV, RWD, and DEV. We discuss the results considering the noPref model with HSMRF-1 prior on $N_{e}(t)$, the adapref models with HSMRF-1 prior on $N_{e}(t)$ and HSMRFs of orders 1 and 2, and GMRFs of orders 1 and 2, on $\beta(t)$. In all cases, we use HMC to estimate the corresponding posterior distribution. Figure \ref{fig:stack+box} top two rows include the boxplots of ENV, RWD, and DEV for the eight simulation scenarios. Each bar in Figure \ref{fig:stack+box} bottom two rows depicts the percentage of simulations each model was one of the top two according to ENV, RWD, and DEV across all simulation scenarios. Each bar refers to a simulation scenario according to one metric. In all plots, we include all three sample sizes considered for each simulation scenario ($1500$ data sets in total). 

Results in Figure \ref{fig:stack+box} confirm that modeling preferential sampling leads to better accuracy. Looking at the bottom two rows, if preferential sampling did not matter, all models would have approximately the same chance of being ranked in the top two ($20$\% each). This is roughly the case for ENV in the CE scenarios. All adaPref models always achieve narrower credible regions (RWD). In terms of DEV, there are a few instances in which ignoring preferential sampling leads to better performance (B-PP, B-LP, B-UP). However, one of the adaPref models (HSMRF-1 prior on both parameters) achieves an identical performance in those scenarios. Figure \ref{fig:stack+box} boxplots also show that the adaPref priors lead to narrower credible regions, and sometimes to smaller mean absolute deviations (DEV).
 
Although we only show results of noPref with HSMR-1 prior on $N_e(t)$ in Figure \ref{fig:stack+box}, we computed the WAIC values for the four MRF priors on $N_{e}(t)$ based on GMRF and HSMRF of orders 1 and 2. The chosen HSMR-1 model achieved the highest WAIC more frequently across the $3000$ datasets generated ($50$ runs, three sample sizes, and four sampling rates for each $N_e(t)$ trajectory).

\section{SARS-CoV-2 in Los Angeles and Santa Clara counties}
\label{sec:covid}
 \begin{figure}[!t]
	\centering
	\includegraphics[scale=0.57]{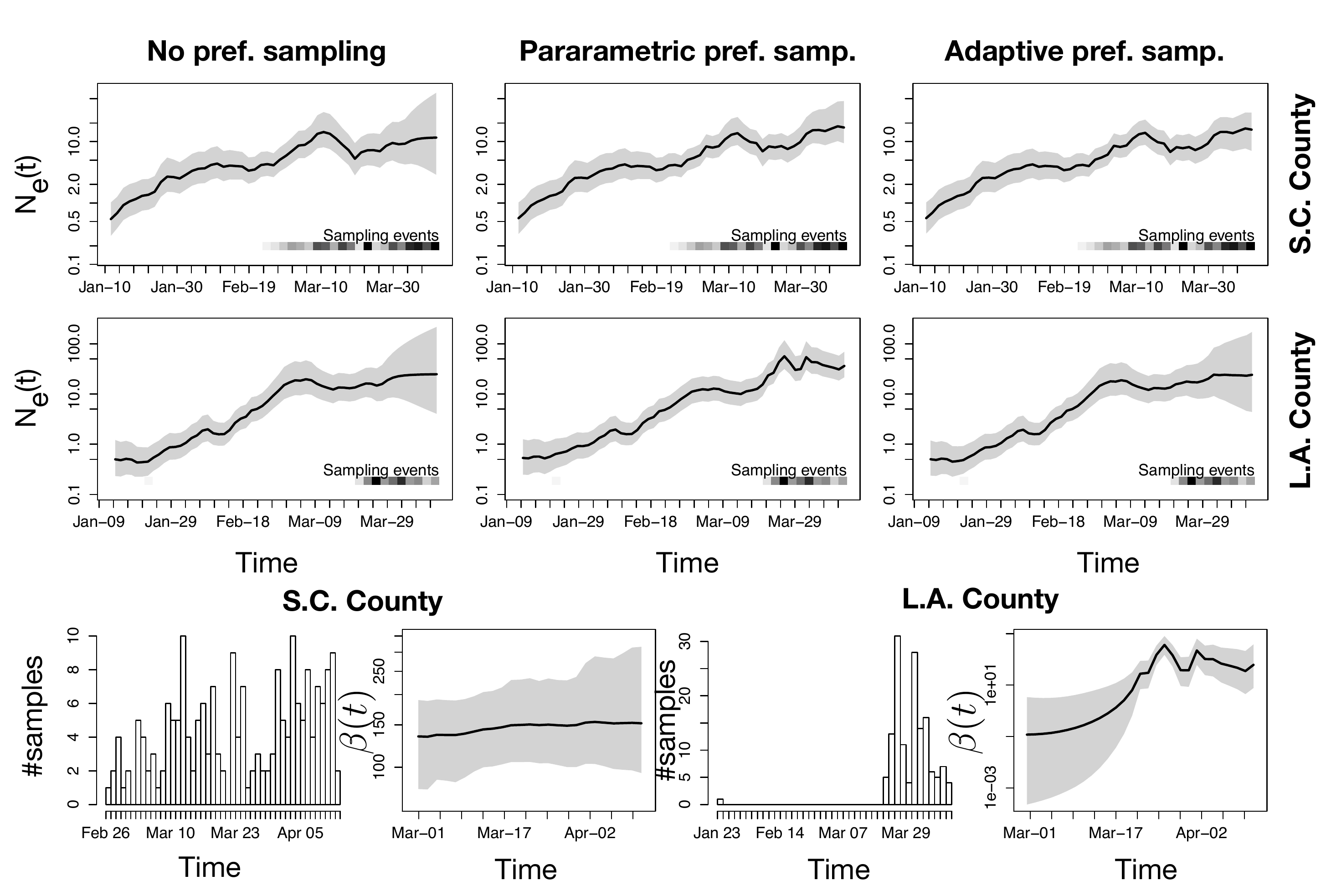}
	\caption{\small{\textbf{{Phylodynamic inference with noPref, parPref and adaPref models from SARS-CoV-2 genealogies inferred from GISAID data obtained from Los Angeles and Santa Clara counties.} }First two rows panels depict $N_e(t)$ posterior distributions inferred using three possible priors: the noPref \citep{pal12}, the parPref \cite{kar16}, and our adaPref prior. The last row first and third panels depict how many samples are collected each day. The second and fourth panels depict $\beta(t)$ posterior distribution (only under the adaPref prior).  The posterior medians are depicted as solid black lines and the 95\% Bayesian credible regions are depicted by shaded areas. Sampling times are also depicted by the heat maps at the bottom of the top two rows panels: the squares along the time axis depicts the sampling time, while the intensity of the black color depicts the number of samples.}}
	\label{fig:covid}
\end{figure}


SARS-CoV-2 is the virus responsible for the coronavirus disease pandemic in 2019-2020. Molecular surveillance of SARS-CoV-2 complements traditional surveillance methods based on case count data and provides a unique opportunity to retrospectively learn past disease dynamics. Here, we use the adaPref model for estimating the viral genetic diversity trajectory $N_{e}(t)$, from currently available viral molecular sequences in GISAID obtained from infected individuals. 
We analyzed viral whole-genome sequences collected in California in Santa Clara (S.C.) and Los Angeles (L.A.) counties and made publicly available in the GISAID EpiCov database \citep{shu2017gisaid}. The GISAID reference numbers of the sequences included in this study, together with data access acknowledgments, are included in the supplementary material.

We downloaded all molecular sequences available on June 27, 2020. The datasets consist of 195 and 134 sequences from S.C and L.A. counties respectively, with collection dates ranging from mid-February, 2020 to April 13 of 2020. 
We included only high coverage sequences with more than 25000 base pairs. 
Sampling frequency information is depicted in the first and third panels of the last row of Figure \ref{fig:covid}. We note that the sampling effort varied in the two counties: most of the L.A. samples are concentrated in late March-mid April, while samples have been collected throughout late February-mid April in S.C. county.

The two estimated genealogies employed in the analysis are the maximum clade credibility trees of the posterior distributions obtained with BEAST2 \citep{bou19}. The MCMC parameters are:  $20 \times 10^6$ iterations, thinning every $1000$ and burnin of $10 \times 10^6$ iterations. We selected the following priors: Extended Bayesian Skyline prior on $N_e(t)$ \citep{hel08}, HKY mutation model with empirically estimated base frequencies \citep{hky}, and uniform prior on the mutation rate with support constrained between $9 \times 10^{-4}$ and $1.1 \times 10^{-3}$ substitutions per site per year. The support of the uniform prior was centered around $1 \times 10^{-3}$ mutations per site per year, an estimate obtained by regressing the Hamming distances of the sequences to the ancestral reference sequence (GenBank MN908947, \cite{wu20}) on the time difference between the sampling times and the reference sampling time.

Given the two estimated genealogies, we approximate posterior marginal distributions of $N_e(t)$ through the INLA approximations of the noPref model \citep{pal12}, the parPref model \citep{kar16}, and our adaPref model. 
In the first two rows of  Figure~\ref{fig:covid}, we show the estimates of effective population size trajectories with the noPref model (first column),  the parPref model (second column), and the adaPref model (third column). Results for S.C. county correspond to the first row and for L.A. county to the second row. Sampling intensity posteriors (computed only through the adaPref model) are given in the third row of Figure \ref{fig:covid} in the second and third panels. 


 The median posteriors of $N_e(t)$ obtained with the noPref and the adaPref models in L.A. county have almost identical trajectories, while the one with the parPref model has a more pronounced maximum later on (around April $1$st). In the S.C. county data set, the median posterior estimates of $N_e(t)$ obtained with the parPref and adaPref models are in this case almost identical, with the estimate obtained with noPref not recovering a steep growth at the end of March. The split behavior of the adaPref posterior, once matching with the noPref posterior and once with parPref posterior, can be explained by looking at the posterior of $\beta(t)$: in the S.C. data set, $\beta(t)$ median posterior is practically flat, a situation consistent with the parametric assumption of the parPref model, while the time-varying $\beta(t)$ accounts for the fact that sampling in L.A. is concentrated in a short time frame. Recall that the parametric assumption of the parPref prior implies a low $N_e(t)$ in February and early March because there are no samples. We believe that this is a positive feature of the adaPref model that it is indeed adaptive to the different sampling protocols.

The average width of the credible regions ($RW=\frac 1 K \sum^{K}_{i=1}|\hat{N}_{97.5}(v_{i})-\hat{N}_{2.5}(v_{i})|$) differ across methods and datasets. In S.C. county, $RW$ is $8.5$ for noPref, $6.6$ for parPref, and $4.7$ for adaPref inferences. In L.A. county, the $RW$ is $23.9$ for noPref, $13.6$ for parPref , and $20.8$ for adaPref inferences. We get a general confirmation that preferential sampling estimators lead to narrower credible regions.

A final remark. The estimates of $N_e(t)$ presented here are representative of genetic diversity over time and do not directly translate to the number of infections. The coalescent we employed ignores recombination, population structure, and selection, which are all assumptions commonly violated in viruses \citep{ram08}. As more scientific knowledge on this virus is produced, the validity of our model assumptions to SARS-CoV-2 will be the subject of further research. Also, we note that observed nucleotide substitutions may be caused by sequencing errors and these are being ignored in our study. 

\section{Discussion}

We have introduced an adaptive preferential sampling model to estimate the effective population size $N_e(t)$ of a coalescent process 
accounting for a situation in which sampling dates are stochastically dependent on the effective population size. We model sampling dates as an inhomogeneous Poisson process with rate $\beta(t) N_e(t)$, where $\beta(t)$ is a time-varying coefficient that modulates how this dependence varies over time. We assume that both {$N_e(t)$ and $\beta(t)$ are continuous functions and model them in a Bayesian framework with Markov random field priors. This methodology allows us to account for preferential sampling while making minimal assumptions on the dependence between the sampling process and the genealogical process. We term the model  proposed adaptive preferential sampling.

The adaptive preferential sampling model allows for a situation in which the sampling protocol changes over time but no detailed knowledge on the way samples are collected is available. In particular, the local adaptivity of the Horseshoe Markov random field prior allows also to model abrupt changes in the sampling protocol. 

We show through simulation studies that the estimates obtained through the adaptive preferential sampling are more accurate than some of the available alternatives, leading to smaller absolute deviations from the true trajectories and narrower credible regions. The performance is competitive also in a broad set of scenarios in which the parametric assumptions of the alternative methods are met. We provide an application to SARS-CoV-2 monitoring and show the ``adaptive nature" of our methodology: in one scenario the estimate was comparable to that of the model without preferential sampling. In a second one, the estimate was matched that of the parametric preferential sampling methodology.

The most direct extension for future work is to include genealogical uncertainty, which is being ignored in the present work. While Kingman heterochronous $n$-coalescent is the standard coalescent model choice to include genealogical uncertainty, recent works have proposed to infer $N_e(t)$ employing lower resolution coalescent models \citep{sai15,cap20}. Our adaptive preferential sampling framework can be paired with any of the ancestral processes.

Another natural extension to the proposed adaptive preferential framework is to incorporate covariates into $\lambda(t)$, the sampling rate, as it is done in \cite{kar20b}, to include auxiliary information about the sampling protocols available to the modeler. For example, it is easy to imagine that one may have direct control over the sampling protocol. The resulting rate of the sampling process would be $\lambda(t)=\beta(t)N_e(t) +\boldsymbol{\beta}'\textbf{X}(t)$, where $\textbf{X}$ is a vector of covariates and $\boldsymbol{\beta}'$ the corresponding linear coefficients. 

In this paper, we model jointly the coalescent process and a sampling process depending on $N_e(t)$. An interesting direction of future work is to model jointly the coalescent process with other processes that depend on $N_e(t)$, such as the total number of infected individuals in an epidemic  \citep{vol09}. The adaptive framework introduced in this paper seems to be suitable to such an extension, given that we make limited assumptions on the dependence between the two processes.

\bibliographystyle{agsm}
\begin{spacing}{1}
	\bibliography{biblio_postdoc}
\end{spacing}

\begin{center}
	{\large\bf SUPPLEMENTARY MATERIAL}
\end{center}

\begin{description}
	
	\item[Observed preferential sampling of SARS-CoV-2 reported in GISAID. ]
	Figure \ref{fig:sampling_dates} depicts the histogram of the sampling dates of the SARS-CoV-2 USA sequences available on GISAID as of  July $20$, $2020$. The red line depicts the daily number of new cases in USA (data downloaded from \texttt{https://github.com/nytimes/covid-19-data}). 
	
	\begin{figure}[!b]
		\centering
		\includegraphics[scale=0.35]{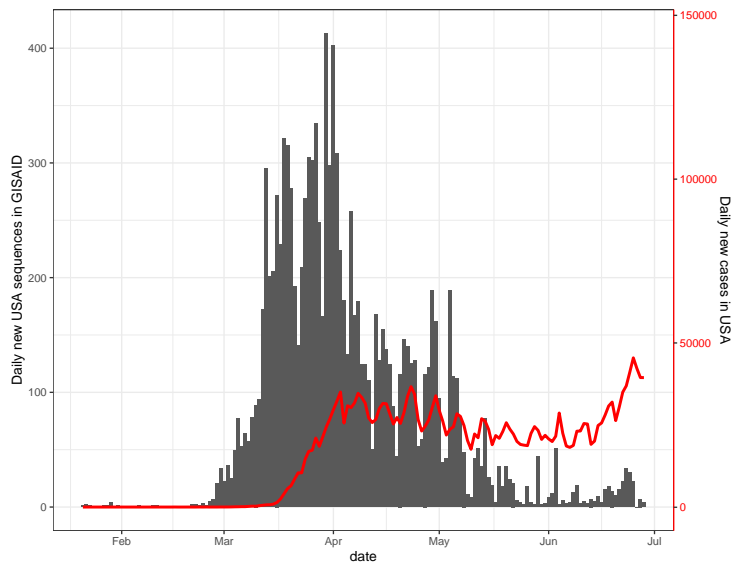}
		\label{fig:sampling_dates}
		\caption{\textbf{Sampling date of SARS-CoV-2 USA sequences available on GISAID as of July $20$, $2020$ and daily new cases in USA in the same period}. }
		
	\end{figure}

	\item[Simulation Details]

	We used simulations to assess the performance of the adaPref priors. We provide here details of the eight simulation trajectories considered (CE and B). In CE scenarios, $N_e(t)$ is  equal to $3$ for $t\leq.15$, $3 \exp(3-20t)$ for $.15<t\leq.3$, and $.15$ otherwise; in B scenarios, $N_e(t)$ is  equal to $.6$ for $t\leq.2$, $.005$ for $.2<t\leq.5$, and $.3$ otherwise. The sampling rate $\lambda(t)$ changes for each simulation scenario. There is a constant of proportionality that will make it change also as a function of the sample size ($n=100,300,500$) in order to ensure that the maximum sampling times is approximately the same.
	
	In CE-PP $\lambda(t)=c_n N_e(t)$ with $c_{n=100}=180$, $c_{n=300}=500$,$c_{n=500}=830$; in CE-U $\lambda(t)=c_n$ with $c_{n=100}=25$, $c_{n=300}=65$,$c_{n=500}=120$; in CE-LP $\lambda(t)=c_n 10 \exp(1.6-20t)$ for $t\leq.23$, and $c_n 0.5$ otherwise, with $c_{n=100}=45$, $c_{n=300}=140$,$c_{n=500}=210$; in CE-UP $\lambda(t)=c_n 10 \exp(3-20t)$ for $t\leq.3$, and $c_n 0.5$ otherwise, with $c_{n=100}=14$, $c_{n=300}=35$,$c_{n=500}=55$. 
	
	In B-PP $\lambda(t)=c_n N_e(t)$ with $c_{n=100}=520$, $c_{n=300}=1700$,$c_{n=500}=3000$; in B-U $\lambda(t)=c_n$ with $c_{n=100}=15$, $c_{n=300}=40$,$c_{n=500}=70$; in B-LP $\lambda(t)=c_n 4$ for $t\leq.1$, $c_n 2$ for $.1<t\leq .4$, and $c_n 4$ otherwise, with $c_{n=100}=40$, $c_{n=300}=130$,$c_{n=500}=200$; in B-UP $\lambda(t)=c_n $ for $t\leq.5$, and $c_n 4$ otherwise, with $c_{n=100}=150$, $c_{n=300}=210$,$c_{n=500}=250$.
	
	\item[Accuracy of adaPref priors in recovering $\beta(t)$.]

	\begin{figure}[!b]
		\centering
		\includegraphics[scale=0.58]{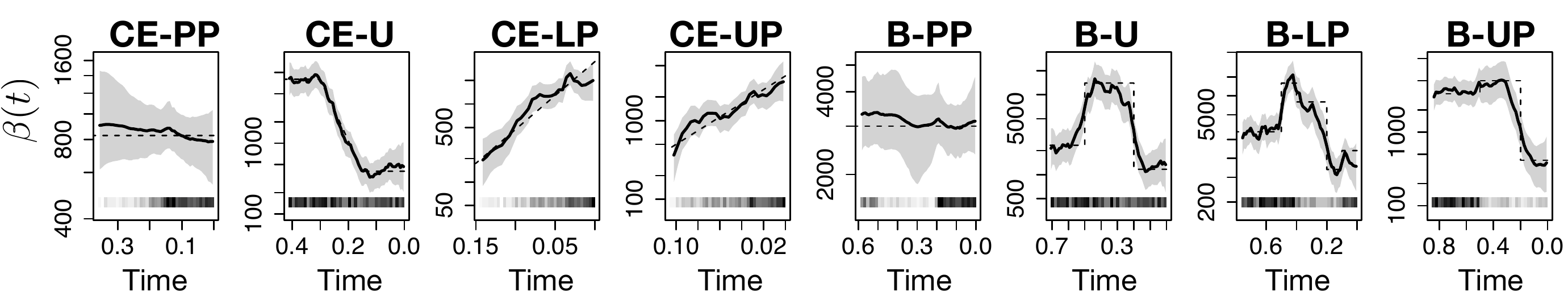}
		\caption{\small{\textbf{Posterior inferences of $\beta(t)$.} Posterior estimates for the eight simulation scenarios and from a single simulated genealogy picked at random with $n=500$ tips. The posterior medians of adaPref are depicted as solid black curves and the 95\% Bayesian credible regions are depicted by shaded areas. \bfn{} and \bfs{} are depicted by the heat maps at the bottom of the last four panels: the squares along the time axis depict the sampling times, while the intensity of the black color depicts the number of samples. True $\beta(t)$ trajectories are depicted as black dashed curves.}}
		\label{fig:traj/beta}
	\end{figure}
	
	\begin{figure}[!b]
		\centering
		\includegraphics[scale=0.6]{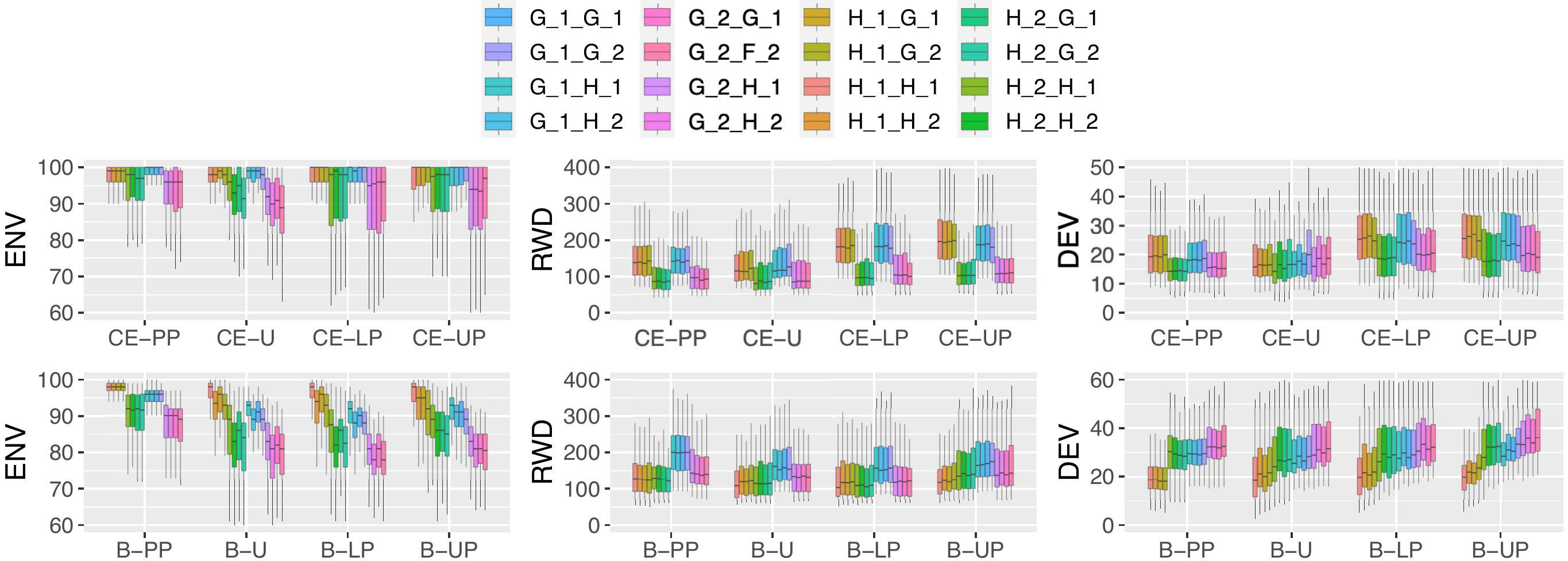}
		\caption{\small{\textbf{{Summary statistics of adaPref posterior inference of $\beta(t)$  grouped by simulation study.} }Each box refers to one method (color in the legend) and depicts the distribution of the $150$ estimated statistics  for each simulation trajectory ($50$ datasets for each sample size, $150$ in total) based on $\beta(t)$ posterior: ENV, first column; RWD, second column; DEV, third column.  In the legend, the first capital letter refers to the type of prior on $N_e(t)$, the second one on $\beta(t)$, with H denoting the horseshoe prior, G the Gaussian prior; the two numbers refer to the corresponding orders.}}
		\label{fig:boxBeta}
	\end{figure}

	We study the ability of the adaPref prior to infer the sampling intensity $\beta(t)$, which we recall to be $\lambda(t)/N_e(t)$.
	Figure \ref{fig:traj/beta} depicts the posterior medians and $95$\% BCI of $\beta(t)$. It shows that $\beta(t)$ is well recovered in all the scenarios. 
	Figure \ref{fig:boxBeta} includes the boxplots of the performance of the $16$ adaPref priors grouped by simulation scenarios. The general conclusion is that empirical accuracy is largely driven by the prior on $N_e(t)$: in CE, models with second-order priors on $N_e(t)$ (both Gaussian and Horseshoe) achieve lower RWD and DEV (boxes in shades of green and purple); in B, models with HSMRF-1 on $N_e(t)$ generally achieve the best performance across the three metrics. We believe that this is a positive feature of the adaPref models because the choice of the prior can be largely based on the prior on $N_e(t)$ which is the actual parameter of interest.

	\item[INLA posterior approximation study]

	\begin{figure}[!b]
		\centering
		\includegraphics[scale=0.55]{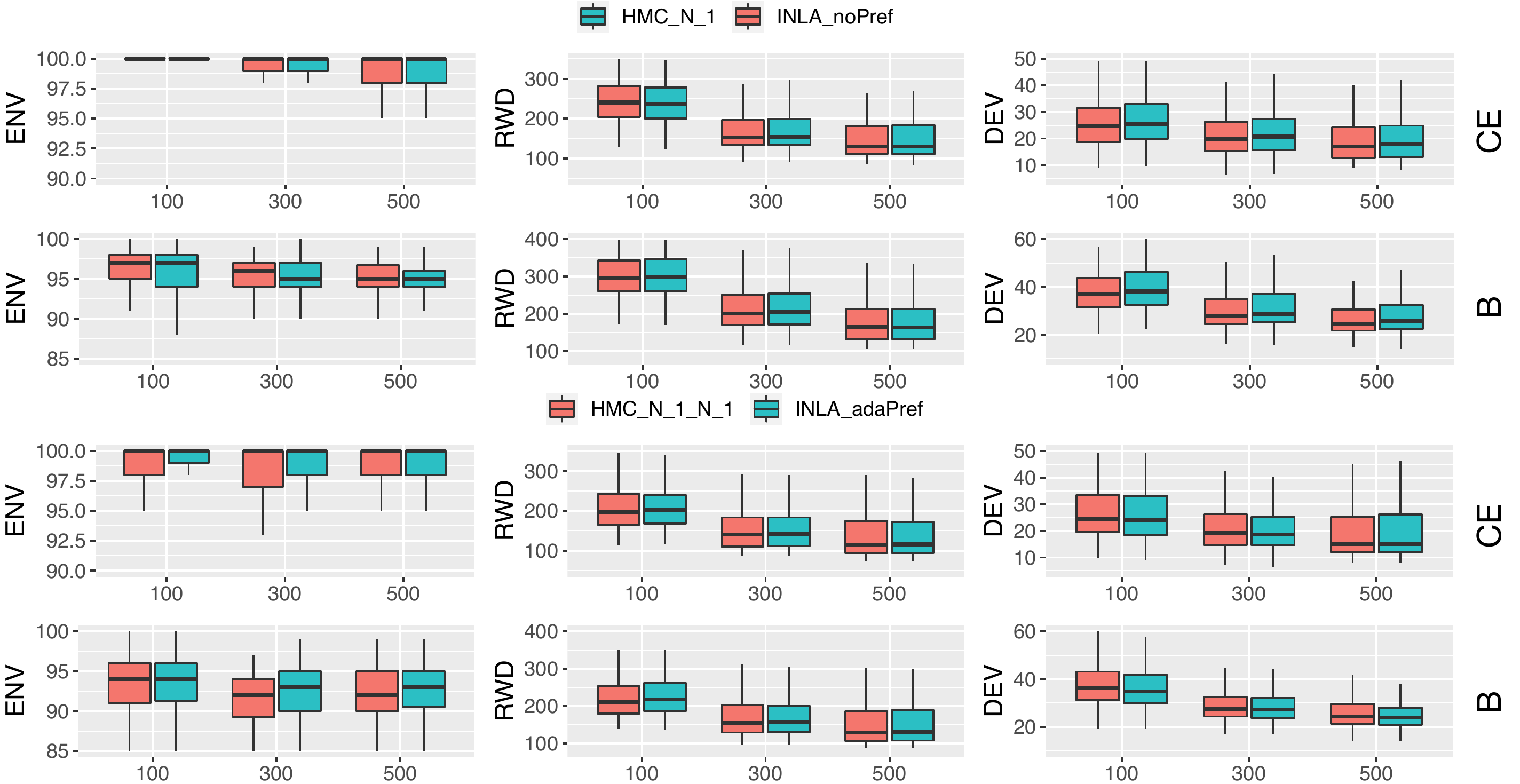}
		\caption{\small{\textbf{Summary statistics of HMC-based and INLA-based posterior inference of $N_e(t)$  grouped by simulation study.} In the first two rows, each box refers to  one method (color in the legend) and depicts the distribution of the $150$ estimated statistics for each simulation trajectory ($50$ datasets for each sample size, $150$ in total) based on $N_e(t)$ posterior: ENV, first column; RWD, second column; DEV, third column. In the third row, the grouping is done according to the sample size: each box is based on $400$ datasets($50$ datasets for each of the eight trajectories). In the legend, HMC\_N\_1\_N\_1 and INLA\_adaPref are our methods, INLA\_noPref is the method in \cite{pal12}, HMC\_N\_1 is the method of \cite{fau20}.}}
		\label{fig:boxINLA_vs_MCMC}
	\end{figure}
	We continue the analysis of the simulation study discussed in Section 4. Here, we study how well INLA \citep{rue09} approximates the posterior marginals of $N_e(t)$ across the different simulation scenarios considered and across different sample sizes ($100,300,500$). In this study we consider the HMC approximated posteriors as the ``ground truth" and see how the INLA approximations perform. We compare the posteriors approximated with INLA to the ones approximated with HMC of the adaptive preferential sampling model with GMRF-1 prios on both $\beta(t)$ and $N_e(t)$. We do an identical comparison for the no preferential model with GMRF-1 prior on $N_e(t)$. 
	
	We check whether the accuracy of INLA approximated posteriors differ from that of HMC approximated posteriors. 
	Once more, we employ ENV, RWD, and DEV to study accuracy. Figure \ref{fig:boxINLA_vs_MCMC} depicts the results grouped by $N_e(t)$ trajectory (CE, B) and sample size. The accuracies under the two approximation methods are largely comparable across simulation scenarios and sample sizes.

	\item[SARS-CoV-2 Molecular Data Description:] Data set used in the applications to SARS-CoV-2. We acknowledge the following sequence submitting laboratories to Gisaid.org: 
	
	\begin{itemize}
		\item \textit{L.A. county data set:} Cedars-Sinai Medical Center, Department of Pathology and Laboratory Medicine, Molecular Pathology Laboratory; California Department of Public Health.
		\item \textit{S.C. county data set:} County of Santa Clara Public Health Department; County of Santa Clara Public Health; Stanford clinical virology lab; Santa Clara County Public Health Department; Chan-Zuckerberg Biohub; Chiu Laboratory, University of California, San Francisco.
	\end{itemize}
	
	We include a description of the sequences accession number and sampling date (location is determined by the county subdivision)
	\begin{itemize}
		\item \textit{L.A. county data set:}{ \small $(EPI\_ISL\_475643 2020-04-13),
			(EPI\_ISL\_475599 2020-04-05),
			(EPI\_ISL\_475617 2020-04-12),
			(EPI\_ISL\_475649 2020-04-03),
			(EPI\_ISL\_475714 2020-04-02),
			(EPI\_ISL\_475603 2020-04-09),
			(EPI\_ISL\_475591 2020-04-11),
			(EPI\_ISL\_467809 2020-03-27),
			(EPI\_ISL\_475607 2020-03-28),
			(EPI\_ISL\_475670 2020-04-01),
			(EPI\_ISL\_475696 2020-03-26),
			(EPI\_ISL\_475684 2020-03-27),
			(EPI\_ISL\_475614 2020-04-06),
			(EPI\_ISL\_475620 2020-04-08),
			(EPI\_ISL\_475679 2020-03-28),
			(EPI\_ISL\_475642 2020-04-01),
			(EPI\_ISL\_475627 2020-03-25),
			(EPI\_ISL\_475629 2020-03-27),
			(EPI\_ISL\_475583 2020-04-02),
			(EPI\_ISL\_475634 2020-03-25),
			(EPI\_ISL\_475577 2020-04-02),
			(EPI\_ISL\_475574 2020-03-27),
			(EPI\_ISL\_475703 2020-03-26),
			(EPI\_ISL\_475652 2020-04-03),
			(EPI\_ISL\_475615 2020-04-12),
			(EPI\_ISL\_475663 2020-04-06),
			(EPI\_ISL\_475666 2020-04-01),
			(EPI\_ISL\_475685 2020-03-26),
			(EPI\_ISL\_475695 2020-03-27),
			(EPI\_ISL\_475654 2020-04-01),
			(EPI\_ISL\_475701 2020-04-04),
			(EPI\_ISL\_475676 2020-04-01),
			(EPI\_ISL\_475651 2020-03-27),
			(EPI\_ISL\_475665 2020-04-06),
			(EPI\_ISL\_475707 2020-03-27),
			(EPI\_ISL\_475677 2020-03-28),
			(EPI\_ISL\_475588 2020-04-02),
			(EPI\_ISL\_475638 2020-03-25),
			(EPI\_ISL\_475624 2020-03-28),
			(EPI\_ISL\_475593 2020-04-04),
			(EPI\_ISL\_475710 2020-04-11),
			(EPI\_ISL\_475630 2020-03-25),
			(EPI\_ISL\_475659 2020-04-09),
			(EPI\_ISL\_475671 2020-04-01),
			(EPI\_ISL\_475669 2020-04-01),
			(EPI\_ISL\_475702 2020-04-04),
			(EPI\_ISL\_475635 2020-03-25),
			(EPI\_ISL\_475705 2020-03-26),
			(EPI\_ISL\_475678 2020-03-28),
			(EPI\_ISL\_475584 2020-03-22),
			(EPI\_ISL\_475664 2020-04-06),
			(EPI\_ISL\_475613 2020-03-26),
			(EPI\_ISL\_475668 2020-04-07),
			(EPI\_ISL\_475590 2020-04-02),
			(EPI\_ISL\_475623 2020-04-08),
			(EPI\_ISL\_475580 2020-04-02),
			(EPI\_ISL\_475657 2020-03-31),
			(EPI\_ISL\_475581 2020-04-02),
			(EPI\_ISL\_475596 2020-03-28),
			(EPI\_ISL\_475694 2020-03-26),
			(EPI\_ISL\_475711 2020-04-11),
			(EPI\_ISL\_475653 2020-03-30),
			(EPI\_ISL\_475633 2020-03-25),
			(EPI\_ISL\_475713 2020-04-02),
			(EPI\_ISL\_475689 2020-03-26),
			(EPI\_ISL\_475661 2020-04-01),
			(EPI\_ISL\_475690 2020-04-03),
			(EPI\_ISL\_475691 2020-04-03),
			(EPI\_ISL\_475609 2020-03-26),
			(EPI\_ISL\_475600 2020-03-26),
			(EPI\_ISL\_475619 2020-04-12),
			(EPI\_ISL\_475640 2020-03-27),
			(EPI\_ISL\_475598 2020-04-05),
			(EPI\_ISL\_475602 2020-04-05),
			(EPI\_ISL\_475660 2020-04-09),
			(EPI\_ISL\_475682 2020-03-28),
			(EPI\_ISL\_475693 2020-03-26),
			(EPI\_ISL\_475639 2020-03-25),
			(EPI\_ISL\_406034 2020-01-23),
			(EPI\_ISL\_475687 2020-03-26),
			(EPI\_ISL\_475700 2020-03-26),
			(EPI\_ISL\_475655 2020-04-01),
			(EPI\_ISL\_475641 2020-03-25),
			(EPI\_ISL\_475674 2020-04-01),
			(EPI\_ISL\_475626 2020-03-28),
			(EPI\_ISL\_475579 2020-04-02),
			(EPI\_ISL\_475625 2020-04-13),
			(EPI\_ISL\_475621 2020-04-06),
			(EPI\_ISL\_475606 2020-04-05),
			(EPI\_ISL\_475611 2020-03-26),
			(EPI\_ISL\_475708 2020-03-27),
			(EPI\_ISL\_475681 2020-03-28),
			(EPI\_ISL\_475618 2020-03-24),
			(EPI\_ISL\_475636 2020-03-25),
			(EPI\_ISL\_475656 2020-03-31),
			(EPI\_ISL\_475622 2020-04-06),
			(EPI\_ISL\_475712 2020-04-02),
			(EPI\_ISL\_475612 2020-03-30),
			(EPI\_ISL\_475686 2020-04-03),
			(EPI\_ISL\_475688 2020-03-26),
			(EPI\_ISL\_475680 2020-03-28),
			(EPI\_ISL\_475644 2020-04-03),
			(EPI\_ISL\_475592 2020-04-11),
			(EPI\_ISL\_475578 2020-03-22),
			(EPI\_ISL\_475715 2020-04-07),
			(EPI\_ISL\_475648 2020-04-03),
			(EPI\_ISL\_475646 2020-04-01),
			(EPI\_ISL\_475587 2020-04-02),
			(EPI\_ISL\_475637 2020-03-26),
			(EPI\_ISL\_475699 2020-03-26),
			(EPI\_ISL\_475704 2020-03-26),
			(EPI\_ISL\_475605 2020-04-09),
			(EPI\_ISL\_475610 2020-03-26),
			(EPI\_ISL\_475582 2020-04-02),
			(EPI\_ISL\_475632 2020-03-25),
			(EPI\_ISL\_475692 2020-03-26),
			(EPI\_ISL\_475716 2020-03-22),
			(EPI\_ISL\_475650 2020-04-03),
			(EPI\_ISL\_475589 2020-03-22),
			(EPI\_ISL\_475586 2020-03-22),
			(EPI\_ISL\_475595 2020-04-05),
			(EPI\_ISL\_475616 2020-04-06),
			(EPI\_ISL\_475594 2020-04-05),
			(EPI\_ISL\_475672 2020-04-13),
			(EPI\_ISL\_475604 2020-04-05),
			(EPI\_ISL\_475628 2020-03-25),
			(EPI\_ISL\_475662 2020-04-13),
			(EPI\_ISL\_475647 2020-04-03),
			(EPI\_ISL\_475576 2020-04-02),
			(EPI\_ISL\_475575 2020-04-02),
			(EPI\_ISL\_475645 2020-04-03),
			(EPI\_ISL\_475673 2020-03-27),
			(EPI\_ISL\_475601 2020-04-05),
			(EPI\_ISL\_475683 2020-03-28),
			(EPI\_ISL\_475706 2020-03-27),
			(EPI\_ISL\_475698 2020-04-07),
			(EPI\_ISL\_475608 2020-03-26),
			(EPI\_ISL\_475585 2020-04-01),
			(EPI\_ISL\_475697 2020-04-04),
			(EPI\_ISL\_475667 2020-04-07),
			(EPI\_ISL\_475658 2020-04-09),
			(EPI\_ISL\_475709 2020-04-02),
			(EPI\_ISL\_475597 2020-04-05),
			(EPI\_ISL\_475631 2020-03-25),
			(EPI\_ISL\_475675 2020-04-01)$}
		
		\item \textit{S.Cla. county data set:}{\small    $( EPI\_ISL\_435599   2020-03-05  ),
			( EPI\_ISL\_435601   2020-03-05  ),
			( EPI\_ISL\_450455   2020-03-24  ),
			( EPI\_ISL\_450456   2020-03-27  ),
			( EPI\_ISL\_437076   2020-04-07  ),
			( EPI\_ISL\_435615   2020-03-07  ),
			( EPI\_ISL\_435627   2020-03-10  ),
			( EPI\_ISL\_476787   2020-03-23  ),
			( EPI\_ISL\_417317   2020-03-02  ),
			( EPI\_ISL\_454668   2020-04-12  ),
			( EPI\_ISL\_454658   2020-04-11  ),
			( EPI\_ISL\_450460   2020-03-24  ),
			( EPI\_ISL\_436647   2020-04-04  ),
			( EPI\_ISL\_476789   2020-03-30  ),
			( EPI\_ISL\_450468   2020-03-18  ),
			( EPI\_ISL\_435654   2020-03-13  ),
			( EPI\_ISL\_476778   2020-03-14  ),
			( EPI\_ISL\_435649   2020-03-13  ),
			( EPI\_ISL\_450458   2020-03-26  ),
			( EPI\_ISL\_435597   2020-03-04  ),
			( EPI\_ISL\_437058   2020-04-06  ),
			( EPI\_ISL\_436646   2020-04-04  ),
			( EPI\_ISL\_435587   2020-02-29  ),
			( EPI\_ISL\_476772   2020-03-23  ),
			( EPI\_ISL\_476791   2020-04-02  ),
			( EPI\_ISL\_417318   2020-02-29  ),
			( EPI\_ISL\_454684   2020-04-12  ),
			( EPI\_ISL\_436672   2020-03-29  ),
			( EPI\_ISL\_450476   2020-03-25  ),
			( EPI\_ISL\_435639   2020-03-12  ),
			( EPI\_ISL\_435653   2020-03-13  ),
			( EPI\_ISL\_435660   2020-03-16  ),
			( EPI\_ISL\_437055   2020-04-06  ),
			( EPI\_ISL\_435590   2020-03-02  ),
			( EPI\_ISL\_476786   2020-03-27  ),
			( EPI\_ISL\_435670   2020-03-19  ),
			( EPI\_ISL\_476780   2020-03-15  ),
			( EPI\_ISL\_437046   2020-04-04  ),
			( EPI\_ISL\_476768   2020-03-25  ),
			( EPI\_ISL\_436659   2020-04-10  ),
			( EPI\_ISL\_450446   2020-03-28  ),
			( EPI\_ISL\_436642   2020-04-01  ),
			( EPI\_ISL\_435657   2020-03-14  ),
			( EPI\_ISL\_450467   2020-03-23  ),
			( EPI\_ISL\_454686   2020-04-12  ),
			( EPI\_ISL\_450445   2020-03-29  ),
			( EPI\_ISL\_454667   2020-04-12  ),
			( EPI\_ISL\_435636   2020-03-11  ),
			( EPI\_ISL\_454679   2020-04-12  ),
			( EPI\_ISL\_454673   2020-04-12  ),
			( EPI\_ISL\_436644   2020-04-02  ),
			( EPI\_ISL\_435580   2020-02-26  ),
			( EPI\_ISL\_435608   2020-03-06  ),
			( EPI\_ISL\_476782   2020-03-13  ),
			( EPI\_ISL\_437049   2020-04-05  ),
			( EPI\_ISL\_450475   2020-03-19  ),
			( EPI\_ISL\_476793   2020-03-18  ),
			( EPI\_ISL\_437064   2020-04-12  ),
			( EPI\_ISL\_435598   2020-03-04  ),
			( EPI\_ISL\_450464   2020-03-23  ),
			( EPI\_ISL\_436677   2020-04-02  ),
			( EPI\_ISL\_435609   2020-03-06  ),
			( EPI\_ISL\_435581   2020-02-28  ),
			( EPI\_ISL\_450474   2020-03-17  ),
			( EPI\_ISL\_435641   2020-03-12  ),
			( EPI\_ISL\_450449   2020-03-25  ),
			( EPI\_ISL\_450466   2020-03-23  ),
			( EPI\_ISL\_450465   2020-03-23  ),
			( EPI\_ISL\_435647   2020-03-13  ),
			( EPI\_ISL\_437052   2020-04-06  ),
			( EPI\_ISL\_436660   2020-04-10  ),
			( EPI\_ISL\_435640   2020-03-12  ),
			( EPI\_ISL\_437057   2020-04-07  ),
			( EPI\_ISL\_435600   2020-03-05  ),
			( EPI\_ISL\_435596   2020-03-04  ),
			( EPI\_ISL\_436654   2020-04-07  ),
			( EPI\_ISL\_437065   2020-04-13  ),
			( EPI\_ISL\_450462   2020-03-23  ),
			( EPI\_ISL\_437085   2020-04-08  ),
			( EPI\_ISL\_450470   2020-03-20  ),
			( EPI\_ISL\_454655   2020-04-11  ),
			( EPI\_ISL\_450479   2020-03-17  ),
			( EPI\_ISL\_454654   2020-04-10  ),
			( EPI\_ISL\_450451   2020-03-25  ),
			( EPI\_ISL\_437045   2020-04-04  ),
			( EPI\_ISL\_429879   2020-03-05  ),
			( EPI\_ISL\_476785   2020-03-11  ),
			( EPI\_ISL\_437043   2020-04-01  ),
			( EPI\_ISL\_435630   2020-03-10  ),
			( EPI\_ISL\_435612   2020-03-07  ),
			( EPI\_ISL\_450477   2020-03-17  ),
			( EPI\_ISL\_450472   2020-03-19  ),
			( EPI\_ISL\_435621   2020-03-09  ),
			( EPI\_ISL\_436661   2020-04-09  ),
			( EPI\_ISL\_435586   2020-03-01  ),
			( EPI\_ISL\_436676   2020-04-01  ),
			( EPI\_ISL\_436667   2020-04-11  ),
			( EPI\_ISL\_450452   2020-03-25  ),
			( EPI\_ISL\_476783   2020-03-13  ),
			( EPI\_ISL\_435671   2020-03-21  ),
			( EPI\_ISL\_437084   2020-04-07  ),
			( EPI\_ISL\_435619   2020-03-10  ),
			( EPI\_ISL\_436665   2020-04-10  ),
			( EPI\_ISL\_435637   2020-03-11  ),
			( EPI\_ISL\_437047   2020-04-03  ),
			( EPI\_ISL\_435631   2020-03-11  ),
			( EPI\_ISL\_436655   2020-04-07  ),
			( EPI\_ISL\_476794   2020-03-30  ),
			( EPI\_ISL\_436656   2020-04-09  ),
			( EPI\_ISL\_435650   2020-03-13  ),
			( EPI\_ISL\_437080   2020-04-04  ),
			( EPI\_ISL\_437048   2020-04-04  ),
			( EPI\_ISL\_417320   2020-03-04  ),
			( EPI\_ISL\_437078   2020-04-07  ),
			( EPI\_ISL\_450473   2020-03-20  ),
			( EPI\_ISL\_437087   2020-04-05  ),
			( EPI\_ISL\_437083   2020-04-09  ),
			( EPI\_ISL\_450480   2020-03-17  ),
			( EPI\_ISL\_436666   2020-04-11  ),
			( EPI\_ISL\_435628   2020-03-10  ),
			( EPI\_ISL\_435582   2020-02-28  ),
			( EPI\_ISL\_476781   2020-03-14  ),
			( EPI\_ISL\_450448   2020-03-28  ),
			( EPI\_ISL\_437050   2020-04-04  ),
			( EPI\_ISL\_450478   2020-03-17  ),
			( EPI\_ISL\_435661   2020-03-16  ),
			( EPI\_ISL\_437077   2020-04-07  ),
			( EPI\_ISL\_476792   2020-04-01  ),
			( EPI\_ISL\_436682   2020-04-08  ),
			( EPI\_ISL\_435594   2020-03-04  ),
			( EPI\_ISL\_476773   2020-03-20  ),
			( EPI\_ISL\_436678   2020-04-03  ),
			( EPI\_ISL\_435618   2020-03-09  ),
			( EPI\_ISL\_436669   2020-04-13  ),
			( EPI\_ISL\_450461   2020-03-23  ),
			( EPI\_ISL\_437051   2020-04-06  ),
			( EPI\_ISL\_436674   2020-03-31  ),
			( EPI\_ISL\_450457   2020-03-21  ),
			( EPI\_ISL\_437059   2020-04-08  ),
			( EPI\_ISL\_436652   2020-04-05  ),
			( EPI\_ISL\_436681   2020-04-03  ),
			( EPI\_ISL\_436645   2020-04-01  ),
			( EPI\_ISL\_436675   2020-04-01  ),
			( EPI\_ISL\_436680   2020-04-04  ),
			( EPI\_ISL\_435638   2020-03-12  ),
			( EPI\_ISL\_435643   2020-03-12  ),
			( EPI\_ISL\_436653   2020-04-09  ),
			( EPI\_ISL\_437056   2020-04-07  ),
			( EPI\_ISL\_437053   2020-04-06  ),
			( EPI\_ISL\_450454   2020-03-24  ),
			( EPI\_ISL\_436648   2020-04-02  ),
			( EPI\_ISL\_437063   2020-04-11  ),
			( EPI\_ISL\_436657   2020-04-09  ),
			( EPI\_ISL\_450459   2020-03-25  ),
			( EPI\_ISL\_437061   2020-04-09  ),
			( EPI\_ISL\_435662   2020-03-16  ),
			( EPI\_ISL\_435658   2020-03-15  ),
			( EPI\_ISL\_454674   2020-04-12  ),
			( EPI\_ISL\_437044   2020-03-31  ),
			( EPI\_ISL\_435645   2020-03-13  ),
			( EPI\_ISL\_436664   2020-04-11  ),
			( EPI\_ISL\_476775   2020-03-19  ),
			( EPI\_ISL\_450469   2020-03-17  ),
			( EPI\_ISL\_435583   2020-02-29  ),
			( EPI\_ISL\_450481   2020-03-18  ),
			( EPI\_ISL\_436663   2020-04-11  ),
			( EPI\_ISL\_450463   2020-03-23  ),
			( EPI\_ISL\_437054   2020-04-05  ),
			( EPI\_ISL\_437086   2020-04-04  ),
			( EPI\_ISL\_436673   2020-04-01  ),
			( EPI\_ISL\_476788   2020-03-14  ),
			( EPI\_ISL\_476779   2020-03-16  ),
			( EPI\_ISL\_435666   2020-03-07  ),
			( EPI\_ISL\_450471   2020-03-19  ),
			( EPI\_ISL\_476784   2020-03-13  ),
			( EPI\_ISL\_435651   2020-03-13  ),
			( EPI\_ISL\_450447   2020-03-28  ),
			( EPI\_ISL\_436643   2020-04-01  ),
			( EPI\_ISL\_454653   2020-04-10  ),
			( EPI\_ISL\_476790   2020-04-02  ),
			( EPI\_ISL\_437088   2020-04-05  ),
			( EPI\_ISL\_450453   2020-03-24  ),
			( EPI\_ISL\_435634   2020-03-10  ),
			( EPI\_ISL\_436679   2020-04-04  ),
			( EPI\_ISL\_436650   2020-04-05  ),
			( EPI\_ISL\_435633   2020-03-11  ),
			( EPI\_ISL\_454670   2020-04-12  ),
			( EPI\_ISL\_436662   2020-04-11  ),
			( EPI\_ISL\_435663   2020-03-16  ),
			( EPI\_ISL\_435667   2020-03-19  ),
			( EPI\_ISL\_476774   2020-03-19  ),
			( EPI\_ISL\_435585   2020-02-29  ),
			( EPI\_ISL\_429880   2020-03-08  ),
			( EPI\_ISL\_436651   2020-04-03  ),
			( EPI\_ISL\_437062   2020-04-10  ),
			( EPI\_ISL\_437060   2020-04-08  ),
			( EPI\_ISL\_435635   2020-03-10  ),
			( EPI\_ISL\_450450   2020-03-25  ),
			( EPI\_ISL\_436641   2020-03-31  ),
			( EPI\_ISL\_436658   2020-04-09  )$}
	\end{itemize}
\end{description}

\end{document}